\documentclass[fleqn,usenatbib]{mnras}

\usepackage{newtxtext,newtxmath}

\usepackage[T1]{fontenc}

\DeclareRobustCommand{\VAN}[3]{#2}
\let\VANthebibliography\thebibliography
\def\thebibliography{\DeclareRobustCommand{\VAN}[3]{##3}\VANthebibliography}
\DeclareRobustCommand{\DE}[3]{#2}
\let\DEthebibliography\thebibliography
\def\thebibliography{\DeclareRobustCommand{\DE}[3]{##3}\DEthebibliography}

\usepackage{graphicx}	
\usepackage{amsmath}	
\usepackage{pdflscape}
\usepackage{rotating}

\PassOptionsToPackage{hyphens}{url}
\usepackage[all]{hypcap}
\usepackage{color}
\definecolor{Brown}{rgb}{0.647,0.165,0.165}
\definecolor{NavyBlue}{rgb}{0.0,0,0.5}
\definecolor{Burgundy}{rgb}{0.5,0.0,0.125}
\hypersetup{
    breaklinks=true,
    colorlinks=true,
    citecolor={Burgundy},
    linkcolor={NavyBlue},
    urlcolor={Brown}
}
\usepackage{multirow}
\def\apj{Astrophys.\ J.}

\def\mnras{Mon.\ Not.\ R.\ Astron.\ Soc.}
\def\aap{Astron.\ Astrophys.}
\def\apjl{Astrophys.\ J.\ Lett.}

\def\physrep{Phys.\ Rep.}
\def\pre{Phys.\ Rev.\ E}
\def\prl{Phys.\ Rev.\ Lett.}

\def\apjs{ApJS}

\def\pasa{PASA}

\def\nat{Nature}
\def\pasp{PASP}

\def\araa{Ann.\ Rev.\ Astron.\ Astrophys.}

\renewcommand{\vec}[1]{\boldsymbol{#1}}	
\newcommand{\dd}{\mathrm{d}}        

\newcommand{\cm}{{\rm cm}}    
\newcommand{\m}{{\rm m}}      
\newcommand{\pc}{{\rm pc}}     
\newcommand{\kpc}{{\rm kpc}}  
\newcommand{\Mpc}{{\rm Mpc}}  

\newcommand{\muG}{\mu{\rm G}} 

\newcommand{\rad}{{\rm rad}}

\newcommand{\brms}{b_{\rm rms}}

\renewcommand{\ne}{n_{\rm e}}

\newcommand{\RM}{\text{RM}}

\newcommand{\npts}{n_{\rm pts}}
\newcommand{\sigRM}{\sigma_{\rm RM}}
\newcommand{\sigRMb}{\sigma_{{\rm RM}_{b}}}
\newcommand{\SF}{\text{SF}} 
\newcommand{\RMSF}{\text{RMSF}} 
\newcommand{\sfn}{{\rm RMSF}^{(n_{\rm pts})}}
\newcommand{\sftwo}{{\rm RMSF}^{(2_{\rm pts})}}

\newcommand{\sfeleven}{{\rm RMSF}^{(11_{\rm pts})}}
\newcommand{\binocoeff}{\mathcal{C}^{\npts-1}_{m}}
\newcommand{\GRF}{{\rm GRF}}
\newcommand{\ngrid}{n_{\rm grid}}
\newcommand{\B}{B_{0}}
\newcommand{\kB}{k_{B_{0}}}
\newcommand{\lB}{\ell_{B_{0}}}
\newcommand{\ellLS}{\ell_{\rm LS}}
\newcommand{\ellLStwo}{\ell_{\rm LS} (2_{\rm pts})}
\newcommand{\ellB}{\ell_{B_{0}}}
\newcommand{\betaLS}{\beta_{\rm LS}}
\newcommand{\kbmin}{k_{b}}
\newcommand{\lb}{\ell_{b}}
\newcommand{\nemean}{{\langle n_{\rm e} \rangle}}
\newcommand{\knemin}{k_{n_{\rm e}}}
\newcommand{\lne}{\ell_{n_{\rm e}}}
\newcommand{\betane}{\beta_{n_{\rm e}}}
\newcommand{\Pb}{{\mathcal{P}(b)}}
\newcommand{\Pne}{{\mathcal{P}(n_{\rm e})}}
\newcommand{\Brat}{B_{0} / b_{\rm rms}}
\newcommand{\LSrat}{2 \ellLS / (2 \kB)^{-1}}
\newcommand{\nptscrit}{n_{\rm pts, crit.}}
\newcommand\Eq[1]{Eq.\,\ref{#1}}
\newcommand\Fig[1]{Fig.~\ref{#1}}
\newcommand\Sec[1]{Sec.~\ref{#1}}
\newcommand\Tab[1]{Table~\ref{#1}}

\newcommand\rev[1]{{#1}}

\title[SF with higher-order stencils]{Structure functions with higher-order stencils as a probe to separate small- and large-scale magnetic fields}

\author[Seta and Federrath]{
Amit Seta \thanks{E-mail:  \href{mailto:amit.seta@anu.edu.au}{amit.seta@anu.edu.au}} and Christoph Federrath
\\
Research School of Astronomy and Astrophysics,  Australian National University, Canberra, ACT 2611, Australia\\
}

\date{Accepted XXX. Received YYY; in original form ZZZ}

\pubyear{2024}

\begin{document}
\label{firstpage}
\pagerange{\pageref{firstpage}--\pageref{lastpage}}
\maketitle

\begin{abstract}
Magnetic fields are an energetically important component of star-formation galaxies, but it is often difficult to measure their properties from observations. One of the complexities stems from the fact that the magnetic fields, especially in spiral galaxies, have a two-scale nature: a large-scale field, coherent over {\rm kpc} scales and a small-scale, random field with a scale of $\lesssim$ $100~{\rm pc}$. Moreover, it is known that the strength of small- and large-scale fields are comparable and this makes it even harder to find their imprints in radio polarisation observations such as the Faraday rotation measure, ${\rm RM}$, which is the integral over the path length of the product of the thermal electron density and the parallel component of the magnetic field to the line of sight. Here, we propose and demonstrate the use of second-order structure functions of ${\rm RM}$ computed with multiple higher-order stencils as a powerful analysis to separate the small- and large-scale magnetic field components. In particular, we provide new methods and calibrations to compute the scale and the strength of the large-scale magnetic field in the presence of small-scale magnetic fluctuations. We then apply the method to find the scale of large-scale magnetic fields in the nearby galaxies M51 and NGC~6946, using archival data and further discuss the need for computing the ${\rm RM}$ structure functions with higher-order stencils. With multiple modern radio polarisation observatories and eventually the Square Kilometre Array, ${\rm RM}$ observations will significantly improve in quantity and quality, and the higher-order stencil structure function techniques developed here can be used to extract information about multiscale magnetic fields in galaxies.
\end{abstract}




\section{Introduction} \label{sec:int}
Magnetic fields play a dynamically important role in galaxies. They influence the star formation process \citep{PadoanN2011, FederrathK2012, KrumholzF2019, PattleEA2023}, control the propagation of cosmic rays \citep{GrenierEA2015, ShukurovEA2017, SetaEA2018, SampsonEA2023, RuszkowskiP2023}, alter the morphology and distribution of gas and gas flows \citep{ShettyO2006, PlanckXXXV_2016, BeattieEA2022}, and may also play a role in the evolution of galaxies \citep{PakmorV2013, VoortEA2021, WhittinghamEA2023}. Despite their significance, estimating magnetic field properties, especially their scales, from observations alone is very difficult and typically requires input from theory, numerical simulations, or additional independent observations. The lack of information on magnetic field properties makes it challenging to fully account for their involvement in star formation and galaxy evolution -- two outstanding problems in modern astrophysics.

Magnetic fields in galaxies are mainly studied through polarimetric observations in radio, submillimeter, and far-infrared wavebands. The primary reasons for difficulties in estimating magnetic fields from these observations are twofold. First, in almost all of these observations, the observable also depends on quantities other than magnetic fields. For instance, the observed synchrotron intensity depends on both the cosmic ray electrons and the magnetic field component perpendicular to the line of sight. Thus, to estimate magnetic fields from synchrotron intensity, assumptions or additional observations about the cosmic ray electron density are required \citep{SetaB2019}. Furthermore, recent simulations \citep{SetaF2022, GentEA2023, DasG2024} and observations \citep{BraccoEA2020, BorlaffEA2021, CampbellEA2022} suggest that magnetic fields vary with the phase of the interstellar medium (ISM), and the associated quantities in the observables also typically exhibit variations with the ISM phase \citep[e.g., cosmic ray propagation varies with the ISM phase \rev{and structure}; see][]{BustardEA2021, ButskyEA2024, ArmillottaEA2024}. These variations add further complexity to existing methods for extracting magnetic fields from observations.

Secondly, these observations provide integrated quantities along the line of sight, and thus knowledge of the path length of emission is necessary to convert these observables to magnetic field properties. Additionally, for observables that depend on the magnetic field component parallel to the line of sight (e.g.,~Zeeman and Faraday rotation measurements), the integration along the path length can lead to cancellation, and thus knowledge about the magnetic field scale is required to estimate even the magnetic field strength. With these caveats in mind, the present work aims to develop a novel technique to extract magnetic field scales in galaxies from observations.

Magnetic fields in galaxies are naturally connected to velocity fields, which, especially in disc galaxies, include both systematic differential rotation and random, turbulent motions. Turbulence in galaxies is driven on a range of scales by a variety of mechanisms \citep{ElmegreenEA2009,KonstandinEA2012,FederrathEA2017}, from stellar feedback on smaller scales \citep[sub-$\pc$ for stellar outflows and $\approx 100~\pc$ for supernova explosions; see][]{MacLowK2004} to gravitational instabilities at larger $\kpc$ scales \citep{KrumholzEA2018}. Broadly, this introduces two-scale magnetic fields: small and large. Moreover, magnetic fields are known to be much weaker in the early Universe  \citep[$\approx 10^{-10}~\muG$; see][]{Subramanian2016} compared to the present-day field strength in galaxies \citep[$\approx 5~\muG$; see][]{Beck2016}. This amplification is explained via a dynamo mechanism, which is the conversion of galaxies' turbulent kinetic energy to magnetic energy \citep{RuzmaikinEA1988, BeckEA1996, BrandenburgS2005, Federrath2016, Rincon2019, ShukurovS2021, Schekochihin2022}. Conventionally, the dynamos are divided into two types: small-scale and large-scale dynamos, each generating fields at a separate range of scales, further emphasising the broad two-scale nature of galactic magnetic fields.

Galactic turbulence amplifies magnetic fields via the turbulent (or fluctuation or small-scale) dynamo, producing random, small-scale (smaller than or comparable to the driving scale of turbulence, usually $\approx 100~\pc$ scale) magnetic fields, and this process has been extensively studied via numerical simulations \citep{HaugenEA2004, SchekochihinEA2004, FederrathEA2011, FederrathEA2014, SetaEA2020, SetaF2021dyn, SetaF2022, KrielEA2022, GentEA2023, SurEA2024}. In disc galaxies, the turbulent dynamo generated small-scale field coupled with density stratification along the disc height ($\alpha$) and differential ($\Omega$) rotation is further amplified and ordered to generate large-scale (usually $\approx$ a few $\kpc$ scales) magnetic fields via the $\alpha$--$\Omega$ (or mean field or large-scale) dynamo\footnote{It is difficult for the large-scale, $\alpha$--$\Omega$ dynamo to operate in irregular and elliptical galaxies because of the lack of significant differential rotation, but the small-scale dynamo can still be active \citep{MossS1996, SetaEA2021} and magnetic fields are observed in such galaxies \citep{MaoEA2008, MaoEA2012, ShahS2021, MahonyEA2022, LivingstonEA2022}.} \citep{RuzmaikinEA1988, BeckEA1996, BrandenburgS2005, Rincon2019, ShukurovS2021}. The large-scale dynamo is seen in some $\kpc$ scale supernova-driven turbulence simulations \citep{GresselEA2008, GentEA2013, BendreEA2015}, but is yet to be observed in full $\approx 30~\kpc$ disc galaxy simulations \citep[see recent reviews by][for numerical challenges in modelling dynamos]{BrandenburgEA2023, Korpi-LaggEA2024}. When all the ingredients required for both types of dynamos (initial weak fields, turbulence, density stratification, and differential rotation) are numerically modelled in simulations, the dynamo-generated magnetic fields show a two-scale nature \citep[e.g.~see fig.~4 and fig.~5 in][]{GentEA2013}. It is usually difficult to separate these two fields, even more so in observations, which is the central problem addressed in this paper.

One of the promising observational probes of magnetic fields in galaxies is the Faraday rotation measure, $\RM$. When linearly polarised light passes through a magnetised plasma, its observed polarisation angle, $\psi_{\rm obs}$, is rotated as a function of wavelength (mostly at radio wavelengths), $\lambda$, as
\begin{align}\label{eq:psi}
\frac{\psi_{\rm obs}}{\rad} = \frac{\psi_{\rm src}}{{\rad}} + \frac{\RM}{\rad~\m^{-2}}~\frac{\lambda^{2}}{\m^{2}},  
\end{align}
where $\psi_{\rm src}$ is the intrinsic polarisation angle of light at the source, and $\RM$ is defined as \citep{KleinF2015, FerriereEA2021}
\begin{align} \label{eq:rm}
\frac{\RM\phantom{0}}{\rad~\m^{-2}} = C~\int_{\rm src}^{\rm obs} \left(\frac{\ne(\ell)}{\cm^{-3}}\right) \left(\frac{\vec{B}(\ell)}{\muG}\right) \cdot \left(\frac{\dd \vec{\ell}}{\pc}\right),
\end{align}
where $C = 0.812$ in units of $\rad ~\m^{-2}~\cm^{3}~\muG^{-1}\pc^{-1}$, $\ne$ is the thermal electron density, $\vec{B}$ is the total magnetic field (vector sum of small- and large-scale components), and $\vec{\ell}$ is the vector distance of the emitting region from the observer, along the line of sight (LOS). Therefore, the scalar product between $\vec{B}$ and $\vec{\ell}$ means that RM is only directly sensitive to the LOS component of the magnetic field. The polarised light can be from a point, external source (such as pulsars, fast radio bursts, and extragalactic radio sources) or an extended, diffuse medium (such as the ISM or the intracluster medium). The $\RM$ can be measured by determining the slope of the observed polarisation angle and the square of the wavelength of light (\Eq{eq:psi}) or, more accurately, using the $\RM$ synthesis technique \citep{Burn1966, BrentjensB2005}. This technique accounts for gaps in the data and also takes into consideration the varying integration limits in \Eq{eq:rm} as a function of the LOS distance in the case of a diffuse medium. The properties of an ensemble of $\RM$s sampling an ensemble of LOSs are affected by both the small- and large-scale magnetic fields, and in this paper, we propose how the second-order structure functions of $\RM$, when computed with multiple higher-order stencils \citep{SetaEA2023}, can differentiate between the small- and large-scale magnetic field components \rev{\citep[the technique could also be applied more generally to a variety of datasets, as described in][]{Cho2019}} and especially helps us to determine the properties of the large-scale field in the presence of small-scale magnetic fluctuations.

The rest of the paper is organised as follows. In \Sec{sec:met} we describe and motivate the numerical model that the new technique is based on. The results from the numerical experiments used to test and calibrate the technique are presented in \Sec{sec:res}. We then apply the new technique to observations of the nearby spiral galaxies M51 \citep{MaoEA2015} and NGC~6946 \citep{Beck2007}, and provide further discussions in \Sec{sec:dis}. Finally, we summarise our work and conclude in \Sec{sec:con}.

\section{Method} \label{sec:met}
We use simple numerical experiments to construct a thermal electron density ($\ne$), as well as small ($\vec{b}$)- and large ($\vec{\B}$)-scale components of magnetic fields with controllable (known) properties. We then generate $\RM$ from them and compute the second-order structure functions of $\RM$, $\sfn$, evaluated with a $n$-point stencil. Finally, the properties of magnetic fields, especially the large-scale field, are extracted from these multi-point second-order structure functions and compared against the known properties of the generated fields to develop a new method/technique that can be applied to observations. In \Sec{sec:cons}, we discuss the method to construct the small- and large-scale magnetic fields and the thermal electron density with given properties, and in \Sec{sec:rmsf}, we describe the method to compute second-order structure functions of $\RM$ with different numbers of points in their stencils. The adopted notation and the range of parameter values used throughout this work are summarised in \Tab{tab:notdef}.

\begin{table*} 
\caption{Description of notation used throughout and range of parameter values.}
\label{tab:notdef}
\begin{tabular}{ll} 
\hline 
Notation & Description [range of values if it is a parameter] \\ 
\hline
$\ngrid^3$ & number of grid points in numerical experiments [$512^3, 256^3, 128^3$] \\
$L$ & side length of the numerical domain \\
$\vec{B}$ & total magnetic field \\
$\vec{\B}$ & large-scale magnetic field  \\
$\B$ & strength of $\vec{\B}$ [$(0.0, 0.03, 0.06, 0.12, 0.25, 0.5, 1.0, 2.0, 4.0, 8.0, 16.0)$ times the strength of the small-scale field] \\
$\kB$ & wavenumber of $\vec{\B}$ [$(2, 4, 8, 16)$ in units of $2\pi/L$] \\
$\lB$ & scale of $\vec{\B}$, $2\pi/\kB$ \\
$\vec{b}$ & small-scale magnetic field \\ 
$\brms$ & root mean square strength of $\vec{b}$ [$(0.0, 0.03, 0.06, 0.12, 0.25, 0.5, 1.0, 2.0, 4.0, 8.0, 16.0)$ times $B_{0}$] \\
$\kbmin$ & minimum wavenumber of $\vec{b}$ [$(32, 64, 128)$ in units of $2\pi/L$]  \\
$\lb$ & maximum ($\sim$ characteristic) scale of $\vec{b}$, $2\pi/\kbmin$ \\
$\Pb$ & power spectrum of $\vec{b}$ from $\kbmin$ to wavenumber $\ngrid/2$ [power law: $k^{0.00}, k^{-1.00}, k^{-1.67}, k^{-2.00}, k^{-3.00}, k^{-4.00}$] \\
$\ne$ & thermal electron density \\ 
$\nemean$ & mean of $\ne$ \\
$\knemin$ & minimum wavenumber of $\ne$ [$(2, 64)$ in units of $2\pi/L$] \\
$\lne$ & maximum ($\sim$ characteristic) scale of $\ne$, $2\pi/\knemin$ \\
$\Pne$ & power spectrum of $\ne$ from $\knemin$ to wavenumber $\ngrid/2$ [power law: $k^{0.00}, k^{-1.00}, k^{-1.67}, k^{-2.00}, k^{-3.00}, k^{-4.00}$] \\
$\RM$ & rotation measure \\
$\SF$ & second-order structure function \\
$\sfn$ & second-order structure function of $\RM$ computed using $n$-point stencil [$\npts=2, 3, 4, 5, 6, 7, 8, 9, 10, 11$] \\
$\sigRM$ & standard deviation of $\RM$ (second saturation) \\
$\sigRMb$ & standard deviation of $\RM$ with $\vec{\B}=0$ (first saturation) \\
$\betaLS$ & slope of $\sfn$ at larger scales (second rise) \\
$\ellLS$ &  extrapolated scale for $\sfn$ at larger scales (second rise) to intersect $\sigRMb$\\
\hline
\end{tabular}
\end{table*}

\subsection{Construction of small- and large-scale magnetic fields and thermal electron density} \label{sec:cons}
On a triply periodic, cartesian, uniform grid with side length $L$, sampled with $\ngrid^3$ points, we define the total magnetic field as
\begin{align} \label{eq:bgen}
\vec{B} &= \vec{B}_{0} + \vec{b} \\ 
&= B_{0} \sin(2 \pi \kB x)~\vec{\hat{z}} +  \vec{\rm GRF} (\brms, \kbmin, \mathcal{P}(b)),
\end{align}
where $\B$ is the strength of the large-scale field along the $z$-direction, $\kB$ is the wavenumber of the large-scale field, $\vec{\GRF}$ implies Gaussian random field vector ($x, y,$ and $z$ components) generated with mean zero, root mean square strength $\brms$, minimum wavenumber ($\sim$ characteristic scale) of the small-scale Gaussian random field denoted by $\kbmin$, a fixed power spectrum, $\mathcal{P}(b)$, and maximum wavenumber always fixed to $\ngrid/2$. The total magnetic field is divergence-free by construction, as the divergence of the mean field is trivially zero, and for the Gaussian random field, the divergence is constructed to be zero via a Helmholtz decomposition where we only keep modes transverse to the local wave vector, using the \texttt{TurbGen} turbulence generator \citep{FederrathEA2010, FederrathEA2021, FederrathEA2022}.

For the thermal electron density, we generate a random lognormal field using a scalar Gaussian random field, $\GRF$, with mean zero, a known standard deviation, $\sigma(\GRF)$, and a given power spectrum, as 
\begin{align} \label{eq:negen}
\ne = \exp\left(\sigma(\GRF)\times\GRF\right).
\end{align}
This operation preserves the spectral features \citep{VioEA2001}, which in turn allows us to control the spectral properties of $\ne$, i.e., the minimum wavenumber, $\knemin$, and power spectrum, $\Pne$. The magnitude of $\ne$ is then finally normalised to have a mean $\nemean$. 

For both the thermal electron density and the small-scale magnetic fields, the power spectra, $\Pne$ and $\Pb$, are assumed to be a power law in wavenumber, $k^{-\alpha}$, with a fixed exponent, $\alpha$, which is varied in the range $\alpha=[0.00, -4.00]$. All wavenumbers, $k$, are in units of $2\pi/L$, such that $k=1$ implies fluctuations on scale $L$, $k=2$ implies fluctuations on scale $L/2$, and so on (this implies that the actual value of $L$ is not important). \rev{For the large-scale field, the minimum $\kB$ is chosen to be $2$ to ensure that the correlation scale of the large-scale magnetic field, and by extension the total magnetic field, is always smaller than the numerical domain, $L$.} Using the generation process \rev{described here}, we can control all the properties of the constructed thermal electron density, as well as the small- and large-scale magnetic fields, with parameters varying over a reasonable range, given in \Tab{tab:notdef}.

\subsection{RM structure functions with higher-order stencils} \label{sec:rmsf}
With the constructed $\ne$ and $\vec{B}$, $\RM$s are computed using \Eq{eq:rm}, assuming a uniform density of emitting sources and that the emission traverses the path length equal to $L$. From the $\RM$ map (a two-dimensional map with $\ngrid^2$ cells), the second-order $\RM$ structure function with $n$-points in the stencil (superscript) is computed as \citep{Cho2019,SetaEA2023}
\begin{align} \label{eq:sfgen}
\sfn(r)  = \left\langle \left\vert\sum_{m=0}^{\npts-1} (-1)^{m}~\binocoeff~\RM\left(\vec{x} + (n_{h} - m) \vec{r} \right)\right\vert^{2} \right\rangle_{\vec{x}},
\end{align}
where $\vec{x}$ is the two-dimensional position vector on the $\RM$ map over which the average, $\langle \rangle_{\vec{x}}$, is computed, $r=|\vec{r}|$ is the separation/scale, $\binocoeff = {\npts-1 \choose m}$ is the binomial coefficient, $n_{h} = \npts/2$ if $\npts$ is even and $= (\npts-1)/2$ if $\npts$ is odd.

At very high values of separation, $r\to\infty$, or more precisely at scales larger than the correlation length of $\RM$ \citep[Eq.~4 in][]{SetaEA2023}, $\sfn$ saturates to
\begin{align} \label{eq:rmsig}
\sfn({r\to\infty}) = \sum_{m=0}^{\npts-1} \left(\binocoeff\right)^2 \sigRM^2,
\end{align}
where $\sigRM$ denotes the standard deviation of $\RM$s.

For the commonly used $2$-point stencil ($\npts=2$), \Eq{eq:sfgen} gives the familiar formula, 
\begin{align} \label{eq:sfgen2}
\sftwo (r) = \left\langle \left\vert\RM (\vec{x} + \vec{r}) - \RM (\vec{x})\right\vert^2 \right\rangle_{\vec{x}},
\end{align}
and \Eq{eq:rmsig} gives the well-known saturation level,
\begin{align} \label{eq:rmsig2}
\sftwo({r\to\infty}) = 2\sigRM^{2}.
\end{align}
Second-order structure functions (SFs) with the $2$-point stencil are widely used in astrophysics and cosmology \citep[see references in][]{Sinha2020} and also in $\RM$ studies \citep{MinterS1996, HaverkornEA2006, Beck2007, HaverkornEA2008, StilEA2011, AndersonEA2015, MaoEA2015, LivingstonEA2021, RaychevaEA2022, LoiEA2022, VanderwoudeEA2024}. 

However, just the $2$-point stencil might not provide accurate results when the region is not well sampled (usually the case with $\RM$ observations) and/or the slopes of the power spectrum of the underlying $\ne$ and $\vec{b}$ are steep. The level of steepness for which the second-order structure function analysis provides accurate results depends on $\npts$, and thus it is advisable to compute multiple structure functions with varying numbers of points per stencil to study the convergence properties of SF \citep[see Sec.~2 and Fig.~5 in][]{SetaEA2023}. This was shown for $\RM$s computed for purely small-scale, Gaussian random magnetic fields and a lognormal $\ne$ in \citet{SetaEA2023}, while here we demonstrate how using multiple structure functions computed with different numbers of points per stencils can disentangle the small- and large-scale magnetic field properties from $\RM$ observations.

We generate $\RM$ maps and compute $\sfn$ with $\npts=2$ to $\npts=11$ for all possible combinations of the parameters listed in \Tab{tab:notdef}. In \Sec{sec:res}, we show a subset of these results, first assuming a constant $\ne$, to show the power of structure functions with higher-order stencils in separating small- and large-scale magnetic fields (\Sec{sec:imprint}), and then describe the method to extract the scale (\Sec{sec:scale}) and strength (\Sec{sec:strength}) of the large-scale magnetic field in the presence of small-scale magnetic fluctuations. Finally, in \Sec{sec:ne} we discuss how the properties of $\ne$ affect the analysis, showing that the results are qualitatively the same. Overall, the results and methods translate well for all parameter combinations in \Tab{tab:notdef}.

\section{Results} \label{sec:res}

\subsection{Imprint of small- and large-scale magnetic fields on second-order $\RM$ structure functions} \label{sec:imprint}

\begin{figure*}
\includegraphics[width=2\columnwidth]{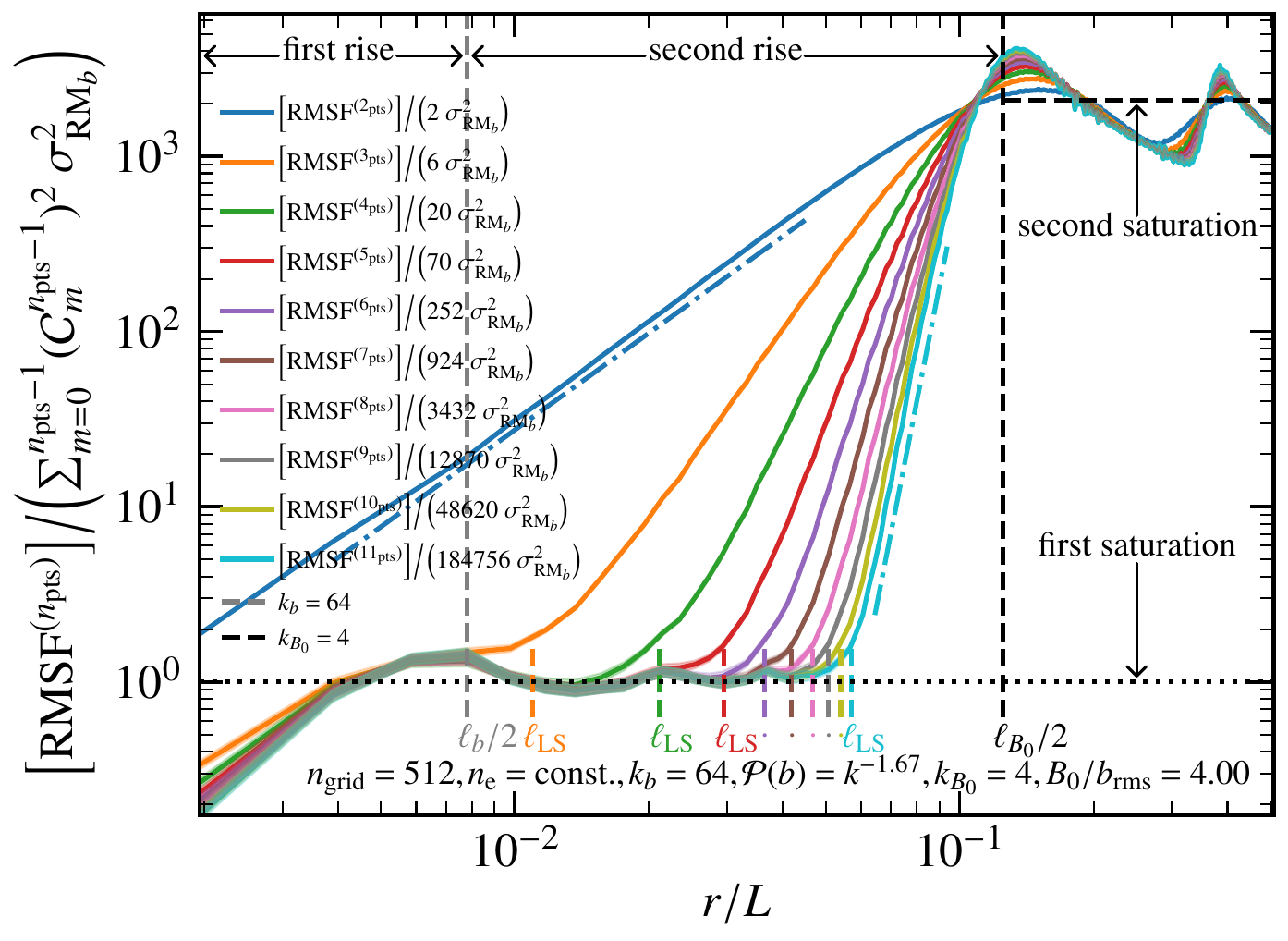}
\caption{Second-order structure function of $\RM$, $\sfn$, computed for a given parameter combination (listed at the bottom of the plot) with $\npts=[2,11]$~number of points per stencils (different colours), normalised by $\sigRMb$ (standard deviation of $\RM$ when the large-scale field, $\B=0$) and normalised by $\sum_{m=0}^{\npts-1} (\binocoeff)^{2}$ from \Eq{eq:rmsig}. Except for the commonly used 2-point stencil SF, all other SFs show signatures of the two-scale magnetic field, i.e., $\sfn$ first rises till $r/L \approx (2 \kbmin)^{-1} = \lb/2$ (dashed grey line), then saturates at $1$ (first saturation), then rises again (second rise) till $r/L \approx (2 \kB)^{-1} = \lB/2$ (dashed black line), and then saturates again (second saturation). The scale range for the first saturation and by extension the two-scale nature shows up better with increasing $\npts$. The second rise (selected based on the $y$-range) is fitted with a straight line in the log-log scale (shown for $\npts=2$ as the dashed-dotted dark blue line, and for $\npts=11$ as the dashed-dotted light blue line). This fitted line is then extrapolated to a $y$-value of $1$ using the fitted slope, $\betaLS$, to obtain $\ellLS$ (shown as the short vertical dashed lines for each $\npts$ in their respective colour, marked $\ellLS$ or $\cdot$). This $\npts$-dependent scale will be used to obtain the scale of the large-scale field in \Sec{sec:scale}.}
\label{fig:allpts}
\end{figure*}

For a set of parameters, we show $\sfn$ for different $\npts$ in \Fig{fig:allpts}, where the $x$-axis ranges from $r/L=1/\ngrid$ (smallest possible separation) to $r/L=1/2$ (maximum distinct separation between points in a triply periodic domain, this factor of $2$ applies to all the input scales and carries throughout the text). For all the cases, the structure functions are first normalised by $\sigRMb$ (standard deviation of $\RM$ with zero mean field), which is independent of $\npts$, and then we apply a $\npts$-dependent normalisation of $\sfn ({r\to\infty})$ (factor $\sum_{m=0}^{\npts-1} (\binocoeff)^{2}$ from \Eq{eq:rmsig}). The usual $2$-point stencil shows that the structure function rises and saturates at a large scale, $r/L \approx (2 \kB)^{-1} = \lB/2$. However, all the SFs with higher-order stencils revel the two-scale nature of the input field, where the structure function first rises with a slope that is determined by the small-scale power spectrum \citep[see Fig.~5 in][]{SetaEA2023}, then saturates around the small-scale field scale, $r/L \approx (2 \kbmin)^{-1}=\lb/2$, at a value of $\approx 1$ ($=\sigRMb$, as per normalisation), then rises again to saturate at the large-scale field scale $r/L \approx (2 \kB)^{-1} = \lB/2$. Furthermore, with a higher number of points in the stencil, the first saturation at the value $1$ between the scales $r/L = \lb/2$ and $r/L = \lB/2$ is for a larger range of scales, with the highest seen for $\npts=11$ in \Fig{fig:allpts}. Thus, higher-order stencils are required to bring out the two-scale nature of magnetic fields from the second-order $\RM$ structure functions. 

Usually from observations, it is difficult to compute $\RM$ structure functions at very small scales (limited by the resolution and number of sources per square degree in the sky) and also at very large scales due to the lack of a significant number of data points at such high separations (e.g.~periphery of a spiral galaxy; furthermore, the detection becomes harder there). The small-scale issue is discussed in \citet{SetaEA2023} and here we primarily discuss the large-scale field question, i.e., we aim to devise a method to find large-scale field properties from the structure functions, showing the two-scale nature. 

First, we fit the `second rise' after $r/L = \lb/2$ with a straight-line on the log-log scale of \Fig{fig:allpts} and obtain the slope as $\betaLS$. Then we extrapolate that line to intersect at $\sfn/\sum_{m=0}^{\npts-1} (\binocoeff)^{2}=\sigRMb$ and obtain the extrapolated scale as $\ellLS$ (this is just an intermediate scale required for the method and may not be physically important). Both quantities, $\betaLS$ and $\ellLS$, depend on $\npts$ and increase with $\npts$. \Fig{fig:allpts} shows that $\ellLS$ (the marked small vertical dashed line) tends to approach a typical scale as $\npts\to\infty$. We model this behaviour to find the scale of the large-scale field in \Sec{sec:scale}.

\begin{figure*}
\includegraphics[width=0.98\columnwidth]{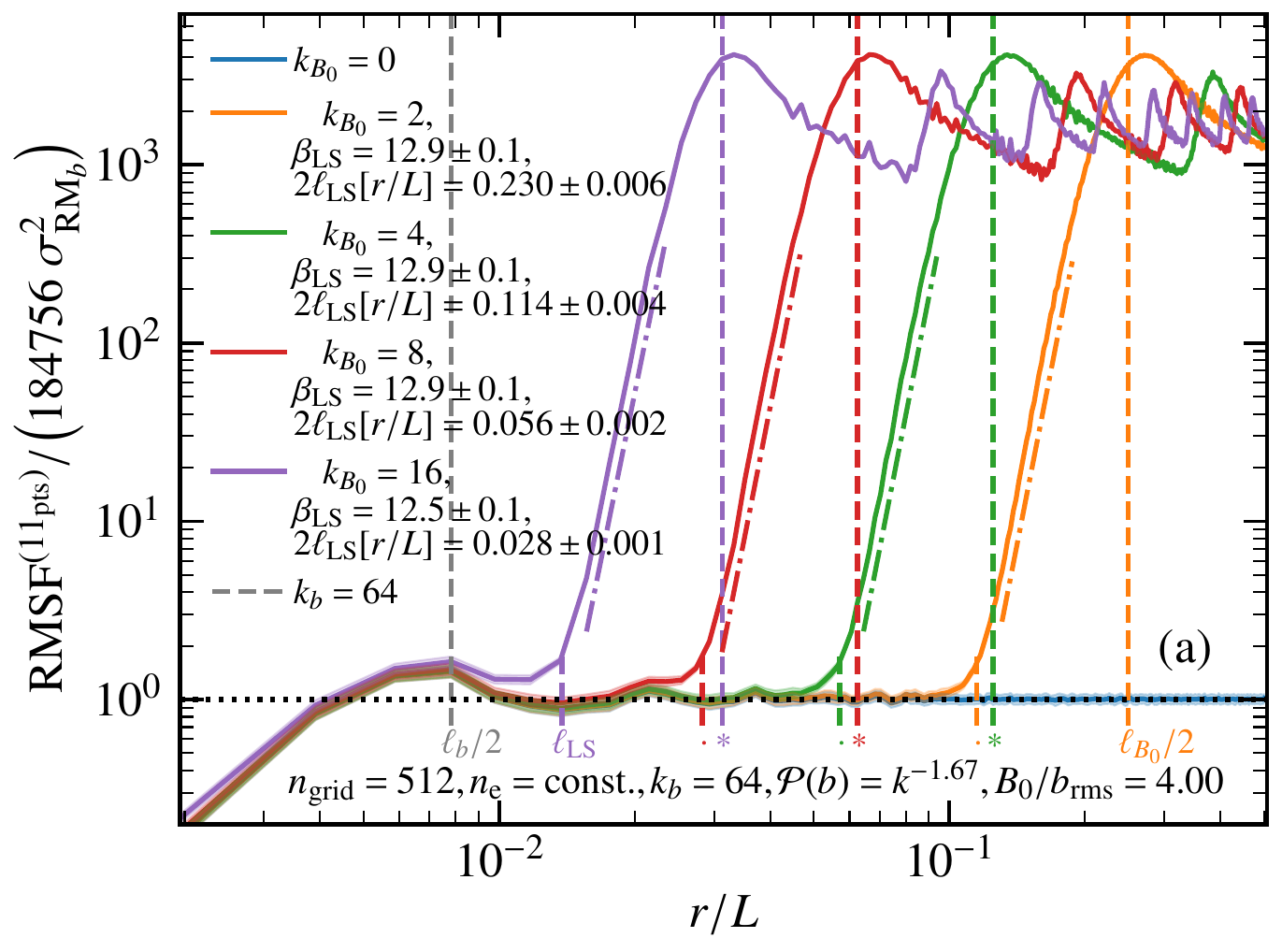}  \hspace{0.5cm}
\includegraphics[width=\columnwidth] {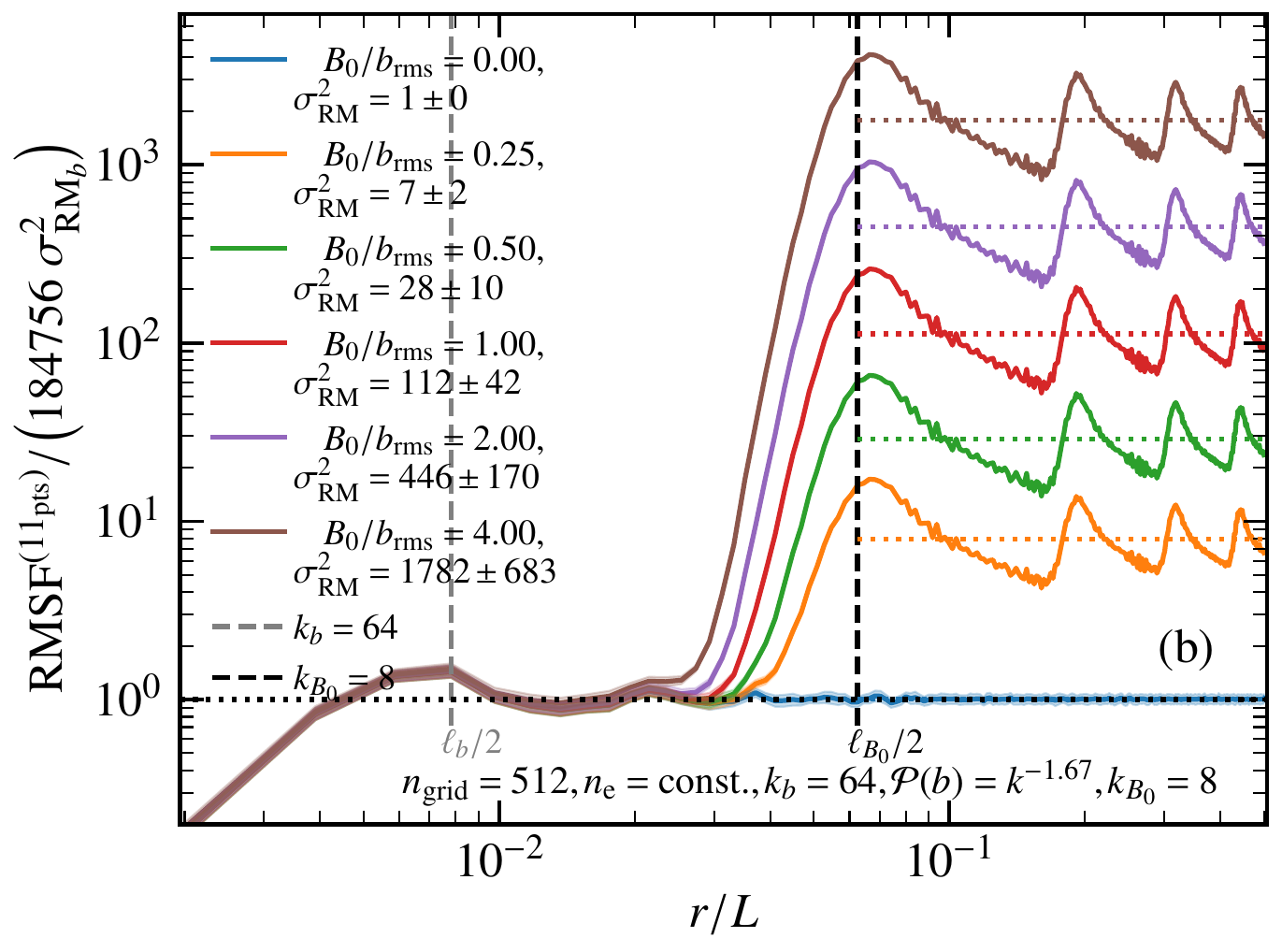}
\caption{The variation of the second-order $\RM$ structure function, $\sfn$, for the highest number of points per stencil, $\npts=11$, with the scale (a) and strength (b) of the large-scale field for otherwise the same parameter combination. (a): With the change in the scale of the large-scale field, $\kB=2,4,8,16$ (long vertical line of different colours, scale marked as $\lB/2$ or $*$), the slope of the second rise (dashed line of respective colours), $\betaLS$, roughly remains the same, but the extrapolated scale, $\ellLS$ (short dashed vertical lines of respective colours, marked as $\ellLS$ or $\cdot$), shifts systematically with $\kB$. (b): Varying the strength of $\B$, the second saturation level \rev{($\sigRM^{2}/\sigRMb^{2}$)} shifts to larger values with increasing $\Brat$. We use these trends to devise a method to derive the scale and strength of the large-scale magnetic field in \Sec{sec:scale} and \Sec{sec:strength}, respectively.}
\label{fig:11pts}
\end{figure*}

Next, in \Fig{fig:11pts}, we show how the normalised $\sfeleven$ (most accurate in the SF series) varies with the scale and strength of the large-scale field. For a fixed ratio of large- to small-scale field strength, $\Brat$, as the scale of the large-scale field increases (or $\kB$ decreases), the level of saturation after the second rise and the slope (dashed-dotted line of respective colour) remains the same, but $\ellLS$ moves towards larger scales in a systematic way as shown in \Fig{fig:11pts}~(a). For a fixed $\kB$, in \Fig{fig:11pts}~(b), we show that when the strength of the large-scale field is increased, the saturation level after the second rise (or $\sigRM$) increases monotonically with $\Brat$. For both the scale and strength variations, the SF values below the scale $r/L = \lb/2$ are unchanged as those are unaffected by the changes in the large-scale field. Also, their first saturation level is always at $\sigRMb$ (for a varying range of scales, depending on the separation between the small- and large-scale field). 

\subsection{Finding the scale of the large-scale magnetic field} \label{sec:scale}

\begin{figure*}
\includegraphics[width=\columnwidth]{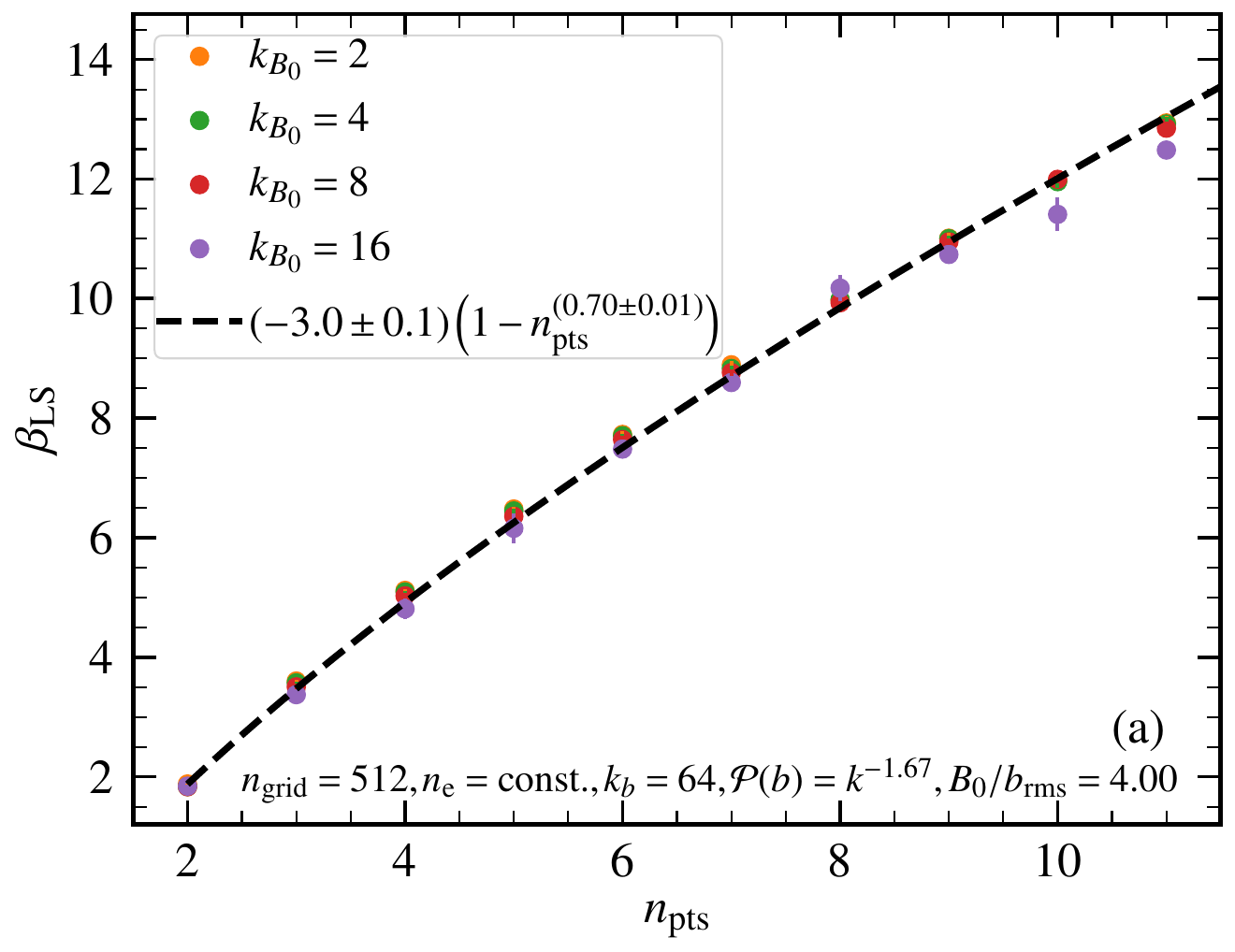}  \hspace{0.5cm}
\includegraphics[width=\columnwidth] {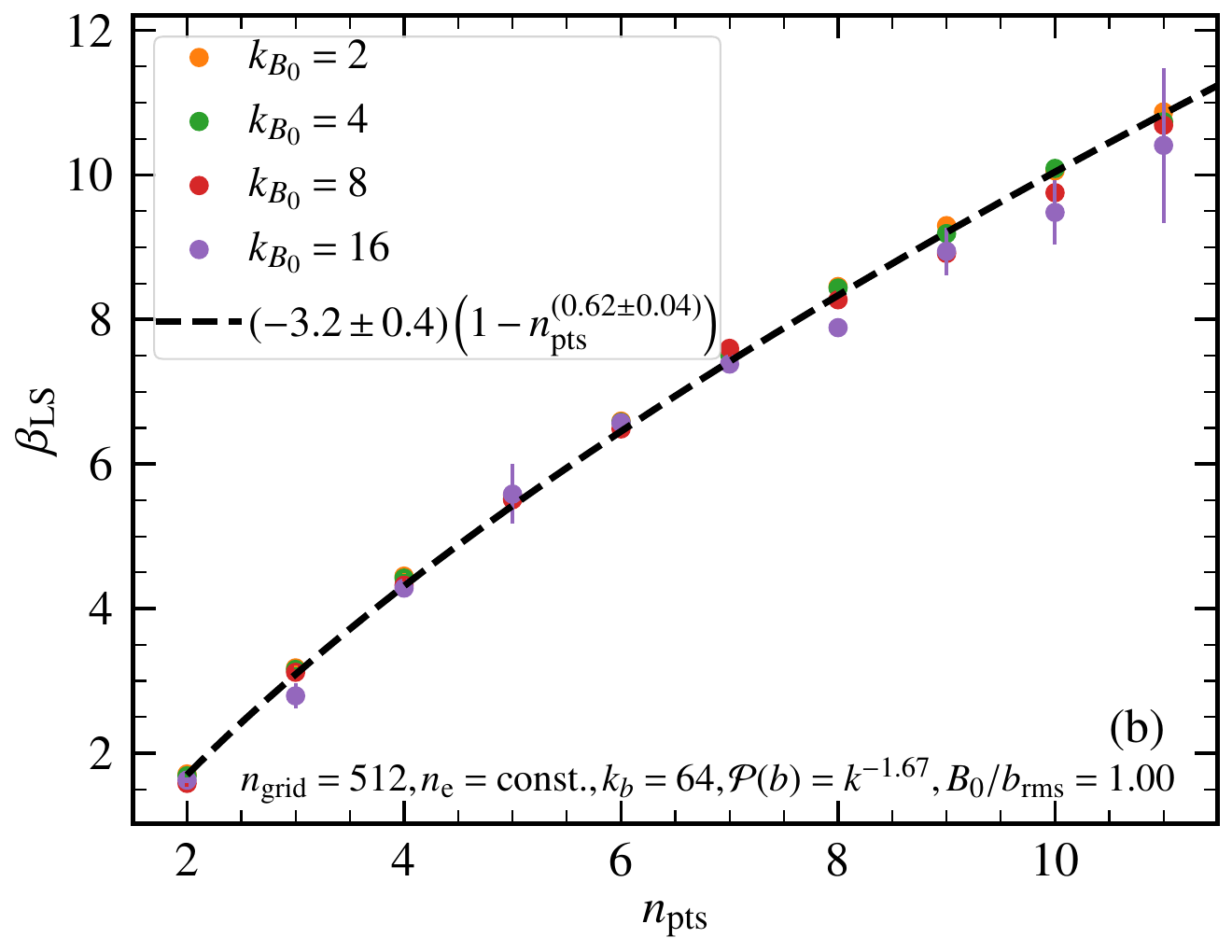}
\caption{Slope of the second rise in $\RMSF$, $\betaLS$, as a function of $\npts$, for different $\kB$ (colours), and two relative field strengths, $\Brat=4$ (a) and $\Brat=1$ (b). The data points are fitted with $a_{\betaLS}(1-\npts^{b_{\betaLS}})$ (dashed black line), showing that $a_{\betaLS}$ and $b_{\betaLS}$ do not vary significantly with $\Brat$.}
\label{fig:slope}
\end{figure*}

\begin{figure*}
\includegraphics[width=\columnwidth]{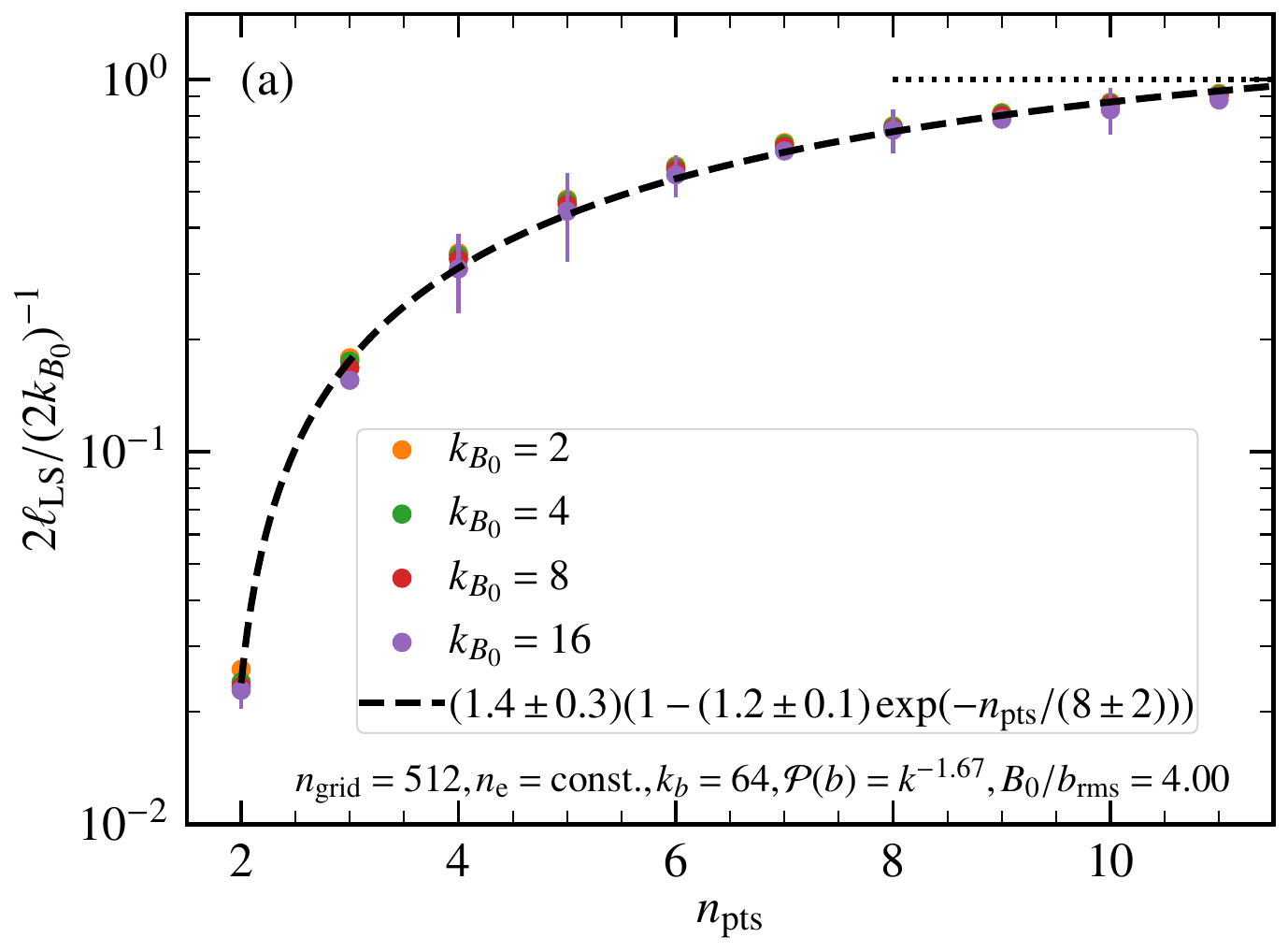}  \hspace{0.5cm}
\includegraphics[width=\columnwidth] {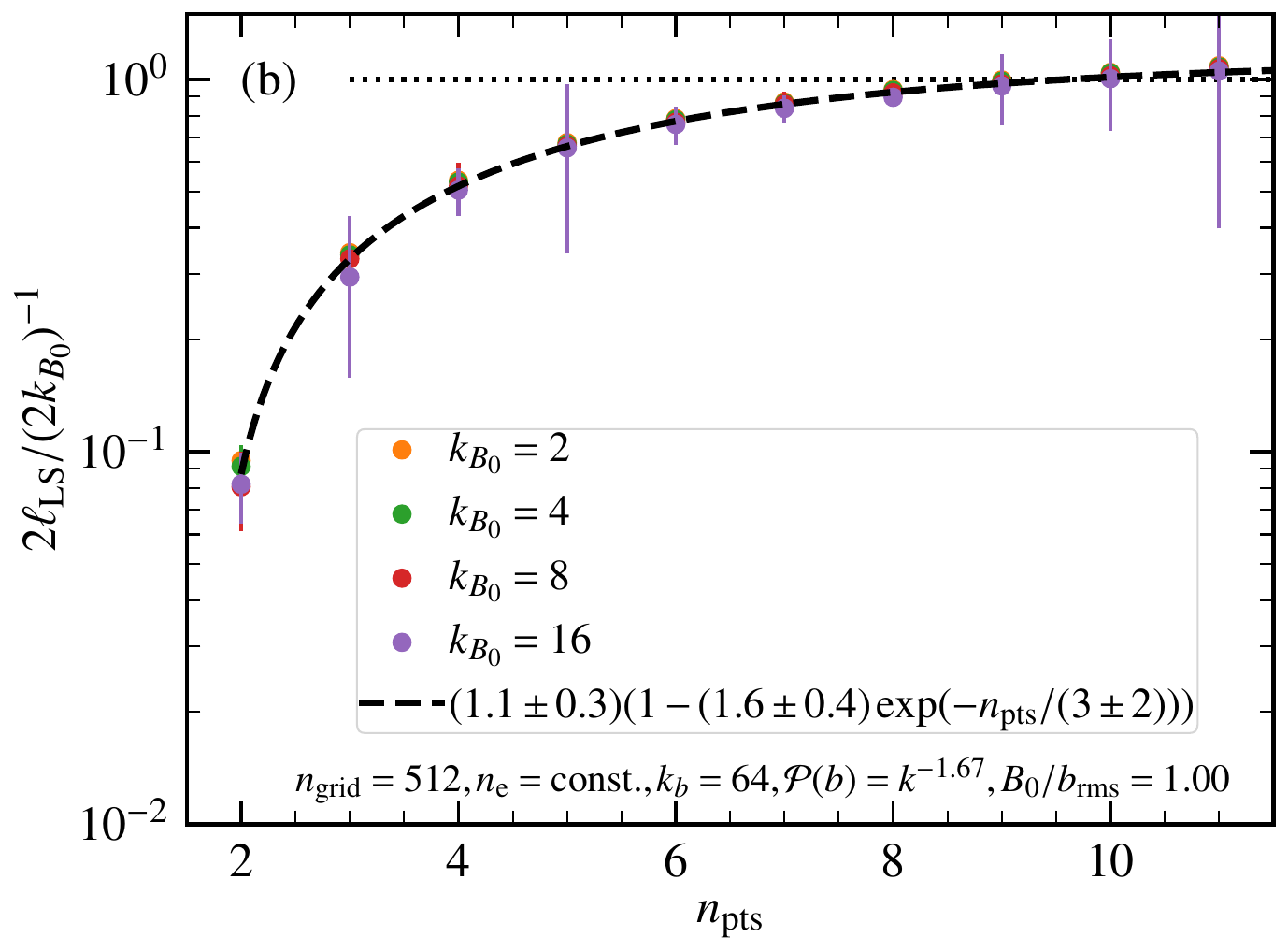}
\includegraphics[width=\columnwidth]{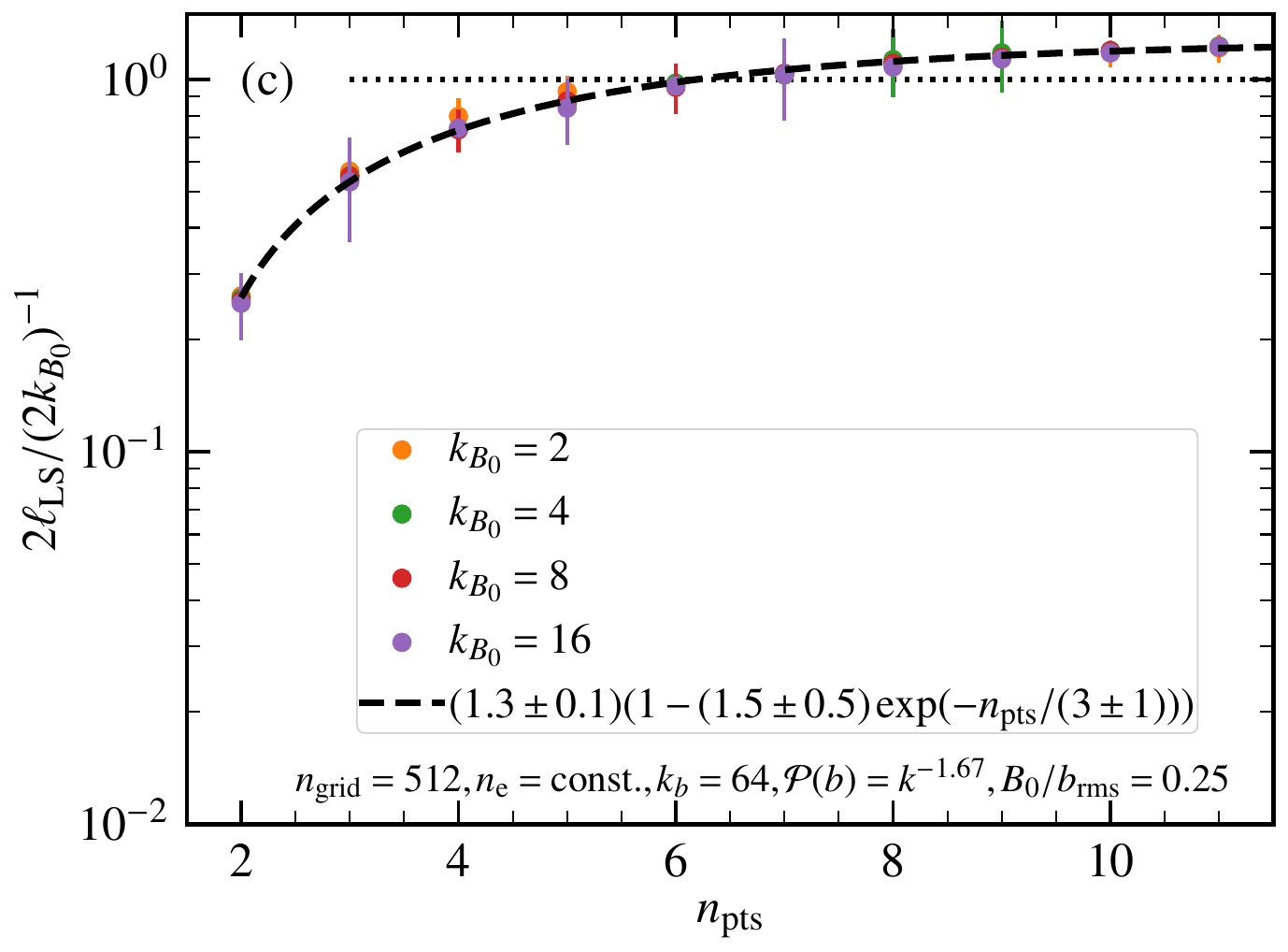}  \hspace{0.5cm}
\includegraphics[width=\columnwidth] {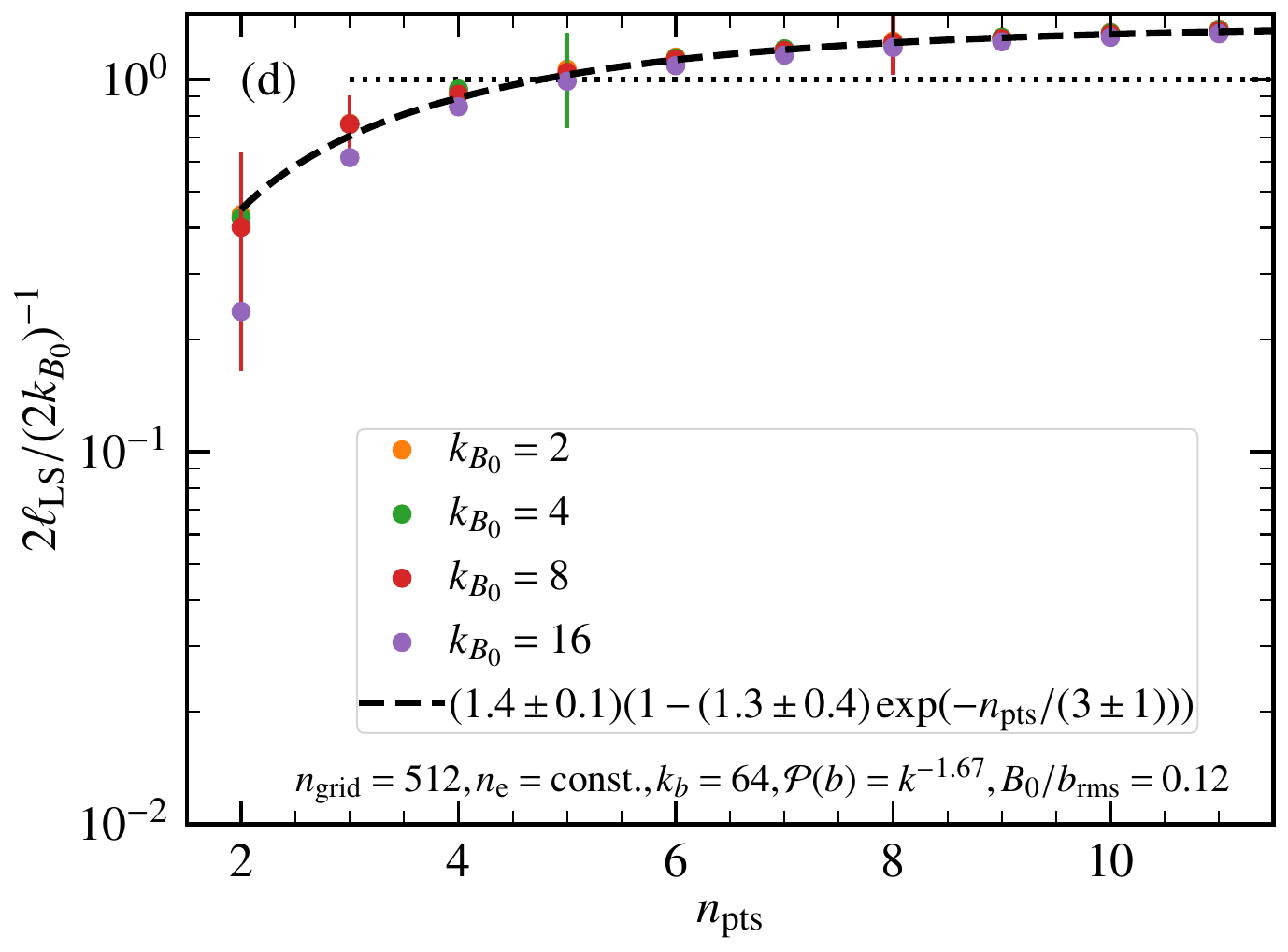}
\caption{Twice the extrapolated scale, $\ellLS$, normalised by half the large-scale field scale, $(2\kB)^{-1} = \lB/2$, as a function of $\npts$, for different $\kB$ (colours), and \rev{four} relative field strengths, \rev{$\Brat=4$~(a), $1$~(b), $0.25$~(c), and $0.12$~(d)}. For all the cases, the ratio $\LSrat$ does not vary much with $\kB$ and approaches $1$ as $\npts\to\infty$. The data fits well with the functional form $a_{\ellLS}~(1-b_{\ellLS}\exp(\npts/\nptscrit))$, with $a_{\ellLS}$ and $b_{\ellLS}$ being statistically independent of $\Brat$. The critical value, $\nptscrit$ depends somewhat on $\Brat$ \rev{for strong large-scale field cases but is obtained to be roughly constant ($\approx 3$) for $\Brat \lesssim 1$}. However, if $\ellLS$ is measured for at least two values of $\npts$, then $\nptscrit$ can be determined independently, and the functional form of $\LSrat$ is fully determined.}
\label{fig:scale}
\end{figure*}

\Fig{fig:slope} shows the slope of the second rise, $\betaLS$, identified in Fig.~\ref{fig:11pts}~(a), for two different field strength ratios, $\Brat=4$~(a) and $\Brat=1$~(b), and varying scale of the large-scale field, $\kB$. In all cases, it follows a functional form of $a_{\betaLS}~(1-\npts^{b_{\betaLS}})$, where $a_{\betaLS}$ and $b_{\betaLS}$ are determined by fitting. We find that the two fit parameters neither vary significantly with the strength nor the scale of the large-scale field. This confirms that the slope of the second rise primarily depends on $\npts$. Thus, the determination of this slope, $\betaLS$, provides a standardisation for the technique.

We now discuss the scale, $\ellLS$, at which the second rise intersects $\sfn / \sum_{m=0}^{\npts-1} (\binocoeff)^{2} = \sigRMb$ (short vertical dashed lines in Figs.~\ref{fig:allpts} and~\ref{fig:11pts}~(a)). \Fig{fig:scale} shows the scale, $\ellLS$, normalised to the large-scale field scale, $\LSrat$, for \rev{$\Brat=4$~(a), $1$~(b), $0.25$~(c), and $0.12$~(d)}, and different $\kB$. We see that as $\npts\to\infty$, $\LSrat\to1$. To determine the number of points required for $\LSrat\to1$, we fit the trend with $a_{\ellLS}[1-b_{\ellLS}\exp(\npts/\nptscrit)]$. We find that $a_{\ellLS}$ and $b_{\ellLS}$ are both $\approx 1$ and do not depend on the scale or the strength of the field. The minimum number of points ($\nptscrit$) required for $\LSrat\to1$ does not significantly depend on $\kB$, but depends on $\Brat$ \rev{for strong large-scale fields, with $\nptscrit=8\pm2$ for $\Brat=4$ but $\nptscrit$ is roughly constant ($\approx 3$) for $\Brat\lesssim1$}. Thus, $\ellLS$ can be fully determined with this relation, by either using a combination of $\ellLS$, $\nptscrit$, and a measurement at a single value of $\npts$, or by using measurements for at least two values of $\npts$, which does not require knowledge of $\nptscrit$. The latter approach is more generic and robust as it does not depend on assuming $\Brat$ (the $\ellLS$ vs.~$\npts$ curve is just extrapolated to $\npts\to\infty$), but the former could be useful when it is not possible to obtain reliable $\RMSF$s with varying $\npts$ (as we do in \Sec{sec:lsobs}).

\subsection{Finding the strength of the large-scale magnetic fields} \label{sec:strength}

\begin{figure}
\includegraphics[width=\columnwidth]{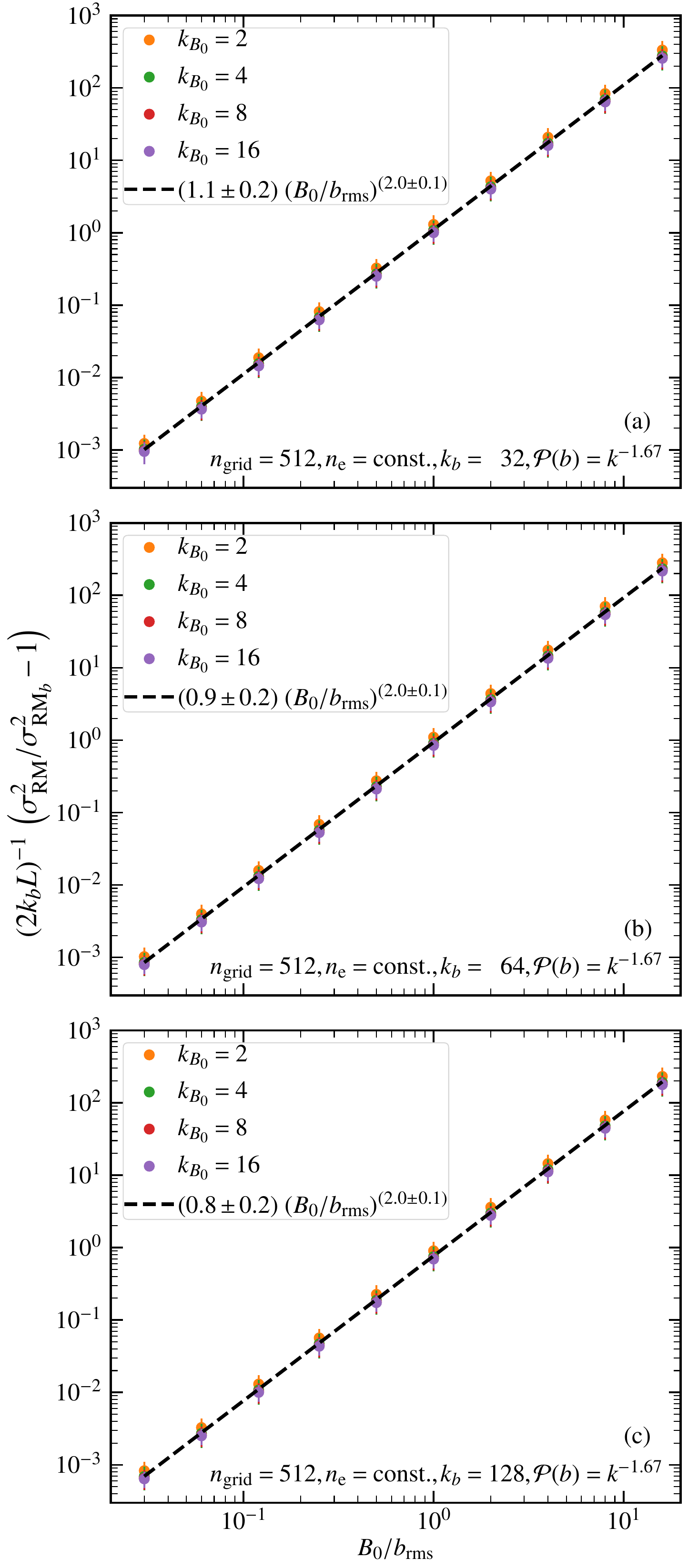} 
\caption{The ratio of large-scale (second) saturation level, $\sigRM$, to the small-scale (first) saturation level, $\sigRMb$, normalised by the product of the small-scale field scale and the size of the domain ($\kbmin L$), averaged over all $\npts$ (whenever the two-scale nature is seen) \rev{with $\npts$-dependent normalisation included in the $\sfn$ (the factor $\sum_{m=0}^{\npts-1} (\binocoeff)^{2}$ from \Eq{eq:rmsig})} as a function of the relative field strength, $\Brat$, for $\kbmin=32$ (a), $64$ (b), and $128$ (c). For all cases, we see $(2 \kbmin L)^{-1}~(\rev{\sigRM^{2}/\sigRMb^{2} - 1}) \approx (\Brat)^2$. Thus, knowing $\kbmin$, $\sigRMb$, and $\sigRM$, one can determine $\Brat$. Then using $\sigRMb$ again, $\brms$ can be determined (\Sec{sec:strength}), and finally, $\B$ can be estimated.}
\label{fig:strength}
\end{figure}

Having determined the scale of the large-scale magnetic field in the previous subsection, we now discuss the method to determine its strength. We saw in \Fig{fig:allpts} that the RMSF shows two saturation (or plateau) regions. First, between the scales $\lb/2$ and $\lB/2$, the plateau gives $\sigRMb$ (standard deviation of $\RM$ for $\B=0$) and the scale range of this plateau is decided based on $\npts$. Second saturation, on scales greater than $\lB/2$, is referred to as $\sigRM$. From \Fig{fig:11pts}~(b) it is clear that the ratio of these two quantities is correlated to the ratio of field strengths, $\Brat$. In \Fig{fig:strength}, we show $(2 \kbmin L)^{-1}~(\rev{\sigRM^{2}/\sigRMb^{2}} - 1)$ as a function of $\Brat$ for all values of $\kB$ and three different values of $\kbmin = 32$~(a), $64$~(b), and $128$~(c). This quantity does not depend on the value of $\kB$ and $\kbmin$ and varies as $\approx (\Brat)^{2}$. Thus, knowing the two saturation regions, i.e., the first $\RMSF$ plateau on smaller scales and the second plateau on larger scales, as well as $\kbmin$ and $L$, we can determine the ratio $\Brat$. Computing $\brms$ from $\sigRMb$ \citep[Eq.~10 in][]{SetaF2021pul}, we can then estimate $B_{0}$. 

In summary, we have described methods to compute the scale (\Sec{sec:scale}) and strength (\Sec{sec:strength}) of the large-scale magnetic fields from the RMSF with higher-order stencils. However, we did so by assuming the thermal electron density, $\ne$, to be constant. We now discuss how $\ne$ properties alter these inferences.

\subsection{Influence of the thermal electron density distribution} \label{sec:ne}

\begin{figure*}
\includegraphics[width=0.98\columnwidth]{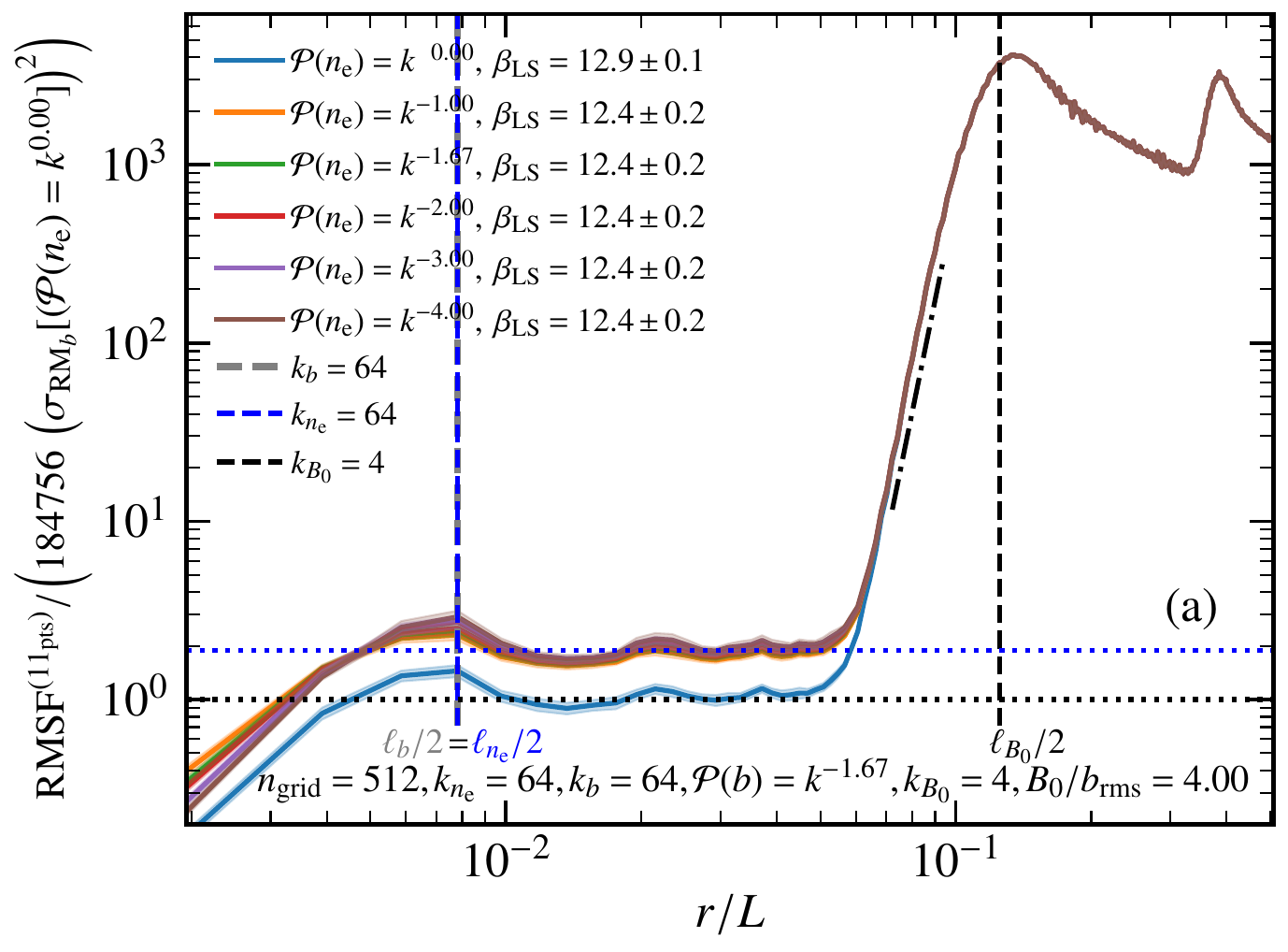}  \hspace{0.5cm}
\includegraphics[width=\columnwidth] {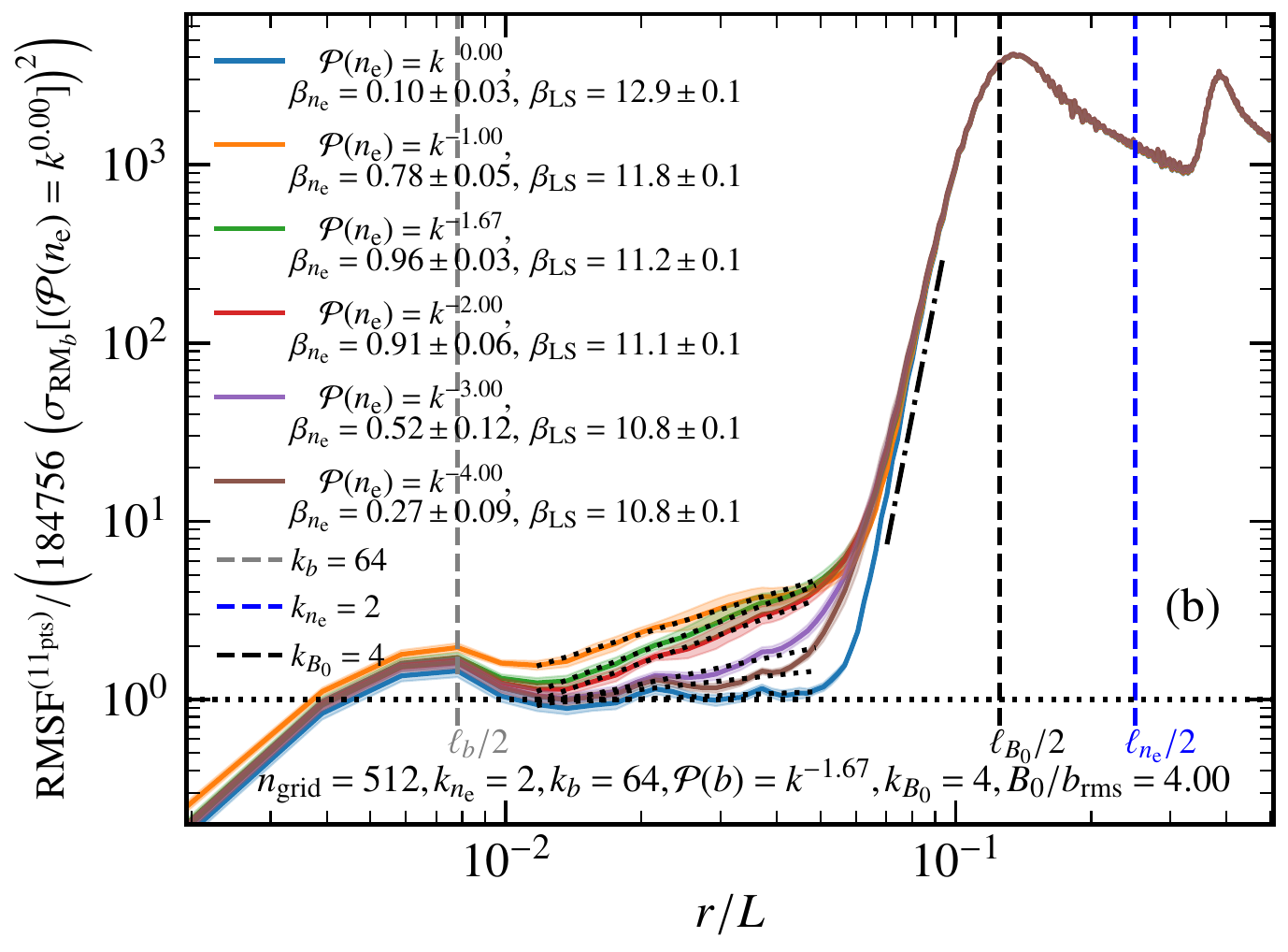}
\caption{$\RMSF$ with an $11$-point stencil, $\sfeleven$, for the parameter combination given in the lower legend of each panel, and the following different properties of the log-normal thermal electron density: power-law slope of the $\ne$ spectrum (different colour) and scale of $\ne$, $\knemin=64$ (small-scale; panel~a) and $\knemin=2$ (large-scale; panel~b). The corresponding scale of the small-scale field, $\lb/2$, the thermal electron density, $\lne/2$, and the large-scale field, $\lB/2$, are shown in grey, blue, and black dashed lines, respectively. For $\knemin=64$, the only difference is the level of first saturation, which shifts to a larger level (dotted blue line) as the variation in $\ne$ also contributes to fluctuations in $\RM$. For $\knemin=2$, the first saturation shows a slope ($\betane$) that depends on the slope of $\Pne$. Thus, this property may be used to constrain the slope of $\Pne$ (given magnetic field properties are known). However, for both cases, the large-scale features in the $\RMSF$, and by extension the methods developed in \Sec{sec:scale} and \Sec{sec:strength}, are not significantly affected by the specific details of the $\ne$ distribution.}
\label{fig:ne}
\end{figure*}

To study the influence of $\ne$ on the methods devised above, we construct a log-normal $\ne$ distribution with a power-law Fourier spectrum, $\Pne$, starting at wavenumber $\knemin$ till $\ngrid/2$, normalised to a value $\langle \ne \rangle$ (see \Eq{eq:negen}). \Fig{fig:ne} shows $\sfeleven$ for various slopes of $\Pne$ and $\knemin = 64$ ($= \kbmin$ or $\lb=\lne$, small-scale, \rev{representing the small-scale clumpy medium in and around HII regions}; panel~a), and $=2$ (large-scale, \rev{representing the large-scale diffuse component}; panel~b). For $\kbmin=64$ (\Fig{fig:ne}~(a)), any non-uniform thermal electron density leads to a change in magnitude of the structure function as fluctuations in $\ne$ also contribute to fluctuations in $\RM$, and thus by extension, the level of $\sigRMb$ increases with $\ne$, but the large-scale features of the $\RMSF$ remain largely unchanged. On the other extreme, when $\knemin=2$ (\Fig{fig:ne}~(b), large-scale $\ne$ fluctuations), between $r/L = (2 \kbmin)^{-1} = \lb/2$ and $r/L = (2 \kB)^{-1} = \lB/2$, we see that $\sfeleven$ no longer remains flat and its slope in that range changes with the slope of $\Pne$ in a systematic way (smaller slope of $\RMSF$ corresponds to a steeper $\Pne$). However, even for this case, the large-scale properties of $\RMSF$ do not depend significantly on the properties of $\ne$ distribution. Furthermore, for these cases, the actual value of $\langle \ne \rangle$ is not important because of the normalisation in computed $\RMSF$s (see $y$-axis in \Fig{fig:ne}).

In galaxies, it is expected that the relevant scale of $\ne$ is likely larger than the scale of the small-scale magnetic fields, but still smaller than the scale of the large-scale fields \citep{GaenslerEA2008, Schnitzeler2012}. In those cases, the structure function in between the scales $r/L = \lb/2$ and $r/L = \lB/2$ would show a mix of both a rising slope on smaller scales (just after $r/L = \lb/2$) like in \Fig{fig:ne}~(b) and a flat region on larger scales (just before $r/L = \lB/2$) like in \Fig{fig:ne}~(a). But in all of those scenarios, the large-scale properties of $\RMSF$ are not largely affected by $\ne$ properties, and the methods to compute the scale and strength of the large-scale magnetic fields in the presence of small-scale magnetic fluctuations described in \Sec{sec:scale} and \Sec{sec:strength}, respectively, are still applicable.

\section{Discussion} \label{sec:dis}
We now apply the methods developed above to existing observational data of second-order $\RM$ structure functions (with 2~points per stencils) for two nearby spiral galaxies, M51 and NGC~6946 (\Sec{sec:lsobs}), and then discuss possible additional effects not covered here in \Sec{sec:mis}.

\subsection{Large-scale magnetic field in M51 and NGC~6946} \label{sec:lsobs}

\begin{figure*}
\includegraphics[width=\columnwidth]{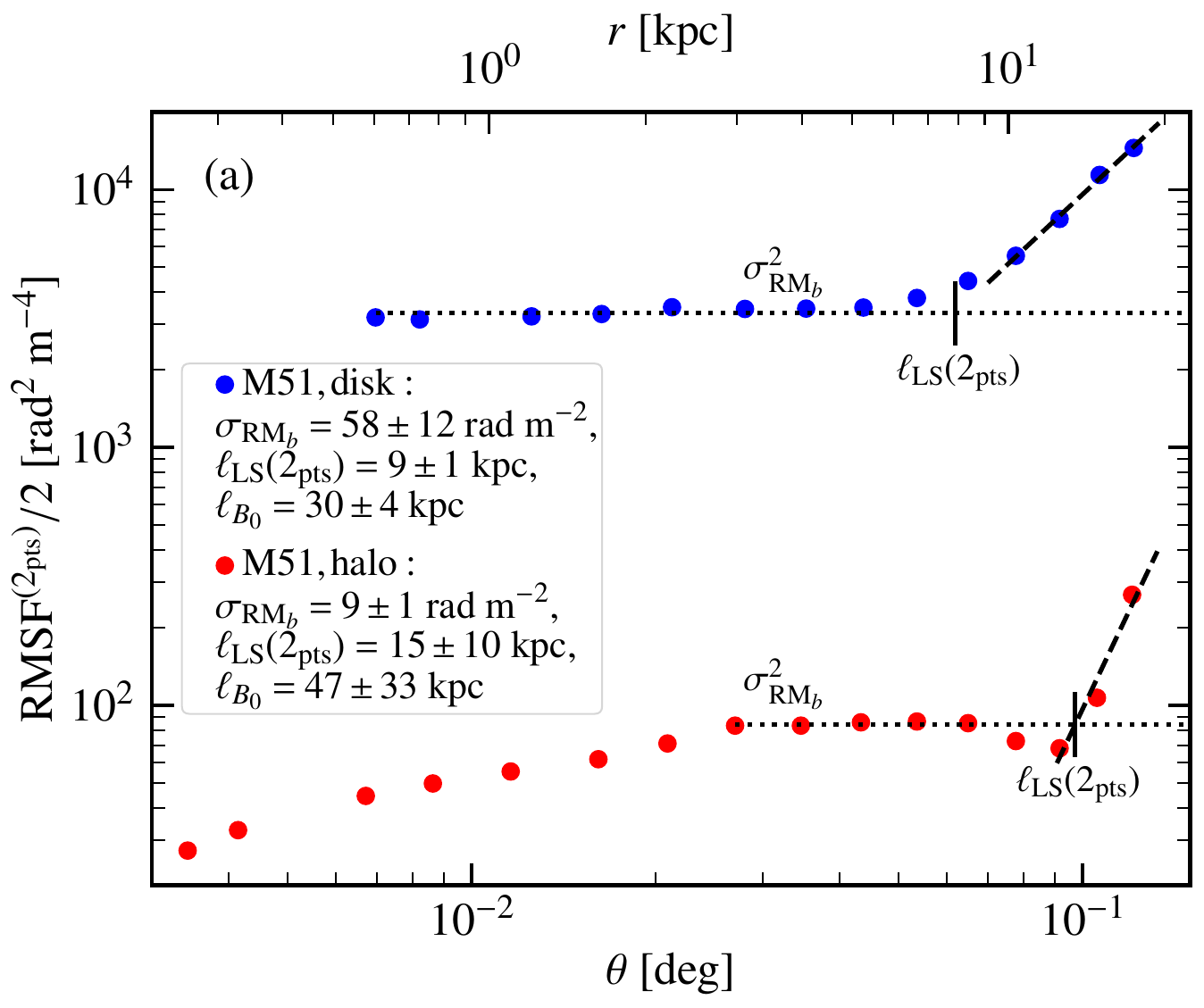} \hspace{0.5cm}
\includegraphics[width=\columnwidth]{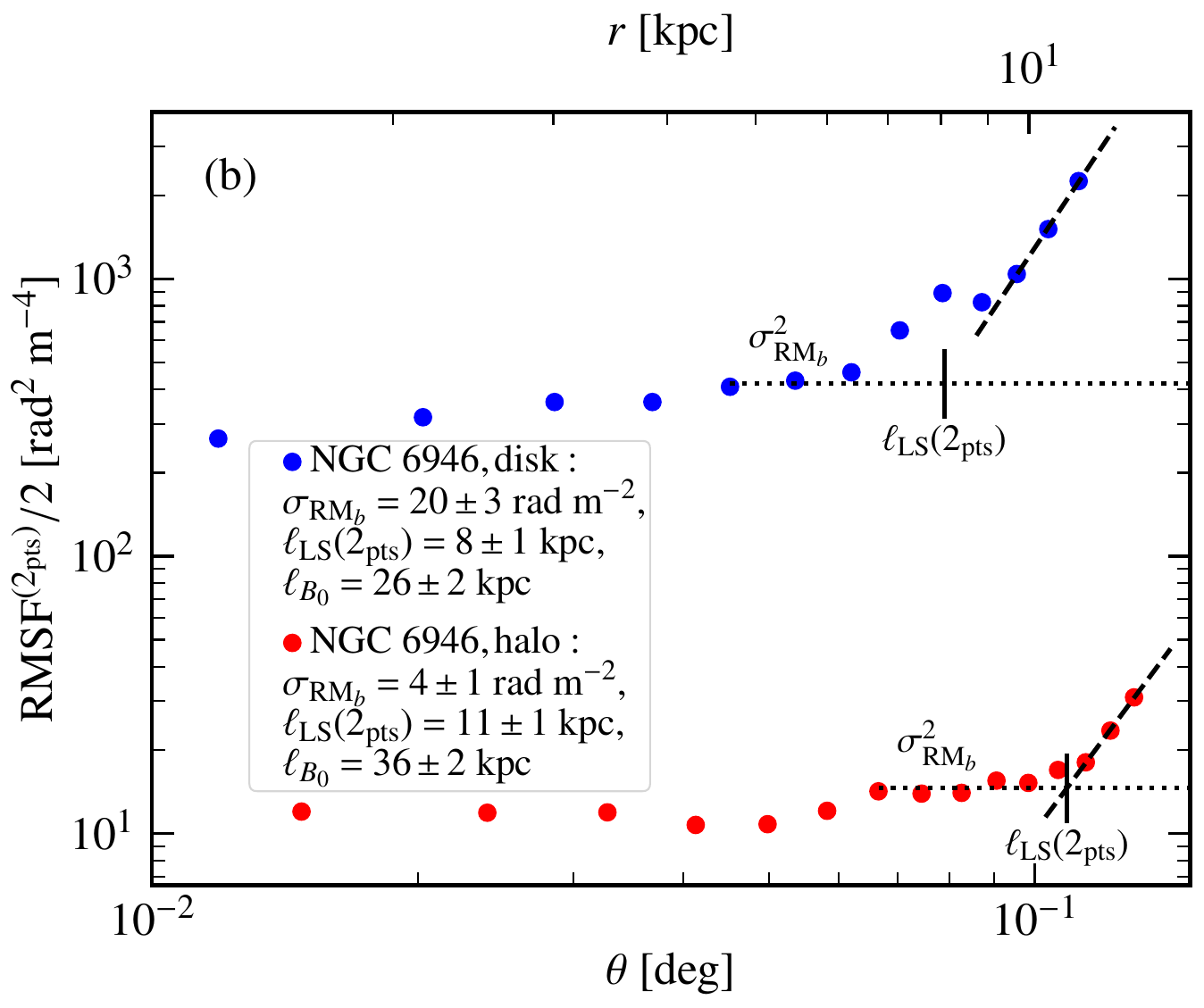}
\caption{The second-order $\RM$ structure function computed with a 2-point stencil, $\sftwo$, for the disk and halo of two nearby, face-on galaxies, M51 and NGC~6946, taken from \citet{MaoEA2015} and \citet{Beck2007}, respectively. In some cases (e.g., for the halo of M51), the $\RMSF$ shows a two-scale nature, i.e., it rises, saturates, and then rises again. However, the large-scale (second) saturation is not obtained for any of these, and thus the strength of the large-scale field cannot be determined from this data. Moreover, the first saturation, $\sigRMb$, (obtained by fitting a flat line to the data before the second rise; dotted black line and marked $\sigRMb^{2}$) and the rise on large scales (obtained by fitting a straight line in log-log space; dashed black line) can be used to obtain $\ellLS$ (vertical solid black line and marked $\ellLStwo$) for all cases. Since from other independent observations and analyses \citep{Beck2007, FletcherEA2011, MaoEA2015, WilliamsEA2024}, it is known that \rev{$\Brat\approx0.1$--$1$} for these galaxies, we can apply \Eq{eq:obsscale} (also see \Fig{fig:scale}~(b, c, and d)), to obtain the scale of the large-scale field, $\ellB$, in the disk and the halo. The scale of the large-scale field in the disk is overestimated and the scale of the large-scale halo field may also be overestimated, at least when compared to existing literature. This might be due to the use of 2~points in the stencils and further demonstrates the need to perform the analysis with multiple $\RMSF$ computed with higher-order stencils, and sensitive data on larger scales, for accurately determining the properties of multi-scale magnetic fields in galaxies.}
\label{fig:obs}
\end{figure*}

We extract the scale of the large-scale magnetic fields in the disk and halo of two nearby, face-on galaxies, M51 and NGC~6946. We use \texttt{WebPlotDigitizer} (\href{https://automeris.io/}{https://automeris.io/}) to extract the second-order RMSF with 2~points per stencil published in \citet{MaoEA2015} for M51 (their fig.~18 for the disk and fig.~19 for the halo), and \citet{Beck2007} for NGC~6946 (their fig.~13). Both datasets are $\sftwo~[\rad^{2}~\m^{-4}]$ as a function of the angular separation on the sky, $\theta~[\deg]$. We convert angular separation into physical separation using a distance of $8.58~\Mpc$ for M51 \citep{McQuinnEA2016} and $5.9~\Mpc$ for NGC~6946 \citep{KarachentsevEA2004}.

\Fig{fig:obs} shows $\sftwo$ (normalised by 2 according to \Eq{eq:rmsig}) as a function of angular ($x$-axis at the bottom) and physical ($x$-axis at the top) separation for both galaxies. The structure functions either show a small-scale rise (slope), saturate, and then the large-scale second rise (slope) or just saturation and then a large-scale rise (in the range $5-10~\kpc$). Thus, it is difficult to ascertain the properties of the small-scale field (e.g., $\kbmin$ and $\Pb$) from this data. Also, since none of the structure-functions shows a second saturation on a larger scale, it is difficult to estimate the strength of the large-scale field using the methods described in \Sec{sec:strength} and \Fig{fig:strength}. However, given the first saturation (used to determine $\sigRMb$) and the second rise on the larger scale is present in all the curves, using the method developed in \Sec{sec:scale}, the scale of the large-scale field, $\ellB = 2 \pi/\kB$, in the disk and halo of both these galaxies can be estimated.

We first determine the plateau/saturation region to estimate $\sigRMb$ for each case. Given, the uncertainties associated with estimating the scale of the small-scale field (see \Sec{sec:scale} for its necessity), the start of the saturation is determined based on fitting a straight line to the data before the large-scale second rise. Then, the saturated value of $\sftwo/2$ is computed and that corresponds to $\sigRMb$. Then the large-scale rise is fitted with a straight line in log-log space and the fitted line is extrapolated to $\sigRMb$ to estimate the scale $\ellLS$. To convert $\ellLS$ to $\ellB$ requires knowing $\Brat$, which cannot be constrained from the data in \Fig{fig:obs} as the $\sftwo$ never saturates a second time on larger scales (as in \Fig{fig:allpts} and \Fig{fig:11pts}). 

From observations and analyses other than $\RM$ structure functions, it is known that the ratio of large- to small-scale magnetic fields, \rev{$\Brat \approx 0.1$--$1$} in both M51 \citep{FletcherEA2011, MaoEA2015, KierdorfEA2020} and NGC~6946 \citep{BeckH1996, Beck2007, WilliamsEA2024}. Thus, we use the equation in \Fig{fig:scale}~(b) \rev{(which is very similar to the equations in \Fig{fig:scale}~(c) and \Fig{fig:scale}~(d) with the same $\nptscrit$, but we use the equation in \Fig{fig:scale}~(b) as it has a higher level of uncertainty in the parameters)} to determine $\lB$ from $\ellLS$ (i.e., $\npts=2$ and the critical number of points for saturation, $\nptscrit = 3$) as
\begin{align}\label{eq:obsscale}
\lB = \frac{4~\ellLS}{2 \pi (1.1 \pm 0.3) (1 - (1.6 \pm 0.4)\exp(-2/(3\pm2))}.
\end{align}
 For M51, we obtain $\ellB = 30 \pm 4~\kpc$ in the disk and $47 \pm 33~\kpc$ in the halo, and for NGC~6946, we find $\ellB = 26 \pm 2~\kpc$ in the disk and $36 \pm 2~\kpc$ in the halo.

For both cases, the scale of the large-scale magnetic field in the disk is a factor of $3$--$6$ times higher than previously discussed for these systems \citep{Beck2007, FletcherEA2011} and more generally for spiral galaxies \citep{Beck2016}. Thus, it is most likely overestimated here. The scale of large-scale magnetic fields in the halo of spiral galaxies is also likely overestimated when compared to results from edge-on spiral galaxies \citep{KrauseEA2018, Mora-PartiarroyoEA2019, KrauseEA2020}. \rev{However, it is known that the circumgalactic medium (CGM; $\sim 100~\kpc$ around the central galaxy) hosts magnetic fields \citep{HeesenEA2023,JungEA2023,RameshEA2023} and if the estimated scale in the halo has contributions from the much larger-scale field in the CGM \citep[large-scale field in the CGM can be due to strong galactic outflows; see][]{PakmorEA2020}, the scale of the large-scale magnetic field might be in the range of values obtained here.} It is important to model and study the scale of the magnetic fields in galaxies, especially in the more diffuse halo/CGM, to account for the impact of magnetic fields on galaxy evolution, and to properly interpret radio polarisation observations for high-redshift galaxies \citep{FarnesEA2014, MalikEA2020, LanP2020, ShahS2021}.

Estimating the scale of magnetic fields from the observed $\RM$s and using the second-order $\RM$ structure function computed from those observations requires further work. First and most essential, it requires computing multiple second-order $\RM$ structure functions with varying number points in their stencils. Here, we only use the available 2-point stencil data for M51 and NGC~6946, but we need to use higher-order stencils ($\npts=3, 4, 5 \cdots$) to allow for a better scale separation (c.f., Fig.~\ref{fig:allpts}), and to remove the necessity of assuming a particular $\Brat$, as with multiple points per stencil, the $\RMSF$ can be extrapolated to $\npts\to\infty$ ($\nptscrit$ determined from the data itself). Second, we need to use a single mathematical model to capture the two-scale nature of $\RMSF$ (e.g.~\Fig{fig:allpts}) \footnote{Even though the discussion in this paper is about two-scale magnetic fields in galaxies, the $\RM$ structure function for a galaxy cluster might also show a two-scale nature \citep{LoiEA2022}, and this analysis is in principle also applicable there.}, possibly four power laws with three breaks, to better estimate its various properties required by the methods described in \Sec{sec:scale} and \Sec{sec:strength}. Third, we need more sensitive data at higher separations (far into the CGM) to capture the second saturation level on larger scales. Here, we merely demonstrate the working principle of the $\RMSF$ methods by using M51 and NGC~6946 data to motivate the need for computing second-order $\RM$ structure functions with higher-order stencils when both large- and small-scale magnetic fields are present. Also, such enhancements might need a larger number of $\RM$ observations with a higher spatial and spectral resolution. These are more readily available now with the Polarisation Sky Survey of the Universe's Magnetism \citep[\href{https://possum-survey.org/}{POSSUM};][]{GaenslerEA2010} using the Australian Square Kilometre Array Pathfinder \citep[\href{https://www.atnf.csiro.au/projects/askap/index.html}{ASKAP};][]{HotanEA2021}, and will improve further with the upcoming Square Kilometre Array \citep[\href{https://www.skao.int/en}{SKA};][]{HealdEA2020}.

\subsection{Missing physical and observational effects} \label{sec:mis}

Throughout the modelling, analysis, and development of the methods to extract the scale and strength of magnetic fields from $\RMSF$, we assumed some simplifications and, here, we discuss them and their possible implications. 

Regarding magnetic field modelling, we assumed a simple one-scale large-scale field and a Gaussian random small-scale field. In principle, the large-scale field itself might be multi-scale in nature, with power distributed over a range of scales \citep{GentEA2013}. In that case, the method might not work if the overlap is large between the small- and large-scale fields, not allowing for a separation. However, if the underlying fields do have sufficient scale separation ($\gtrsim 2$--$4$), the methods developed here should work.

We assumed a sinusoidal form for the large-scale field (the results remain the same when a saw-tooth form is used), but the large-scale field might have a more complicated functional form, possibly due to the intricacies of the large-scale dynamo and also as discussed with several empirical large-scale magnetic field models \citep{ShukurovEA2019, BeckEA2019, MaEA2020}. The small-scale field is assumed to be a Gaussian random field without coherent structures, but it can be non-Gaussian with spatially intermittent, filamentary structures as suggested by numerical magnetohydrodynamic simulations \citep{SetaEA2020, SetaF2021dyn} and hinted by filaments seen in recent observations \citep{ZaroubiEA2015, TuricEA2021}. However, for a large ratio of path length to the scale of the random field, the small-scale structure might not matter much for the $\RM$ due to the central limit theorem (assuming thermal electron density and magnetic fields are uncorrelated). 

For the $\RM$ computation, we assumed that magnetic fields and thermal electron densities are uncorrelated. If they are correlated, magnetic field strengths estimated using $\RM$s can be overestimated by a factor of $2$--$3$ \citep{BeckEA2003}. However, this may only be important for smaller (usually sub-$\kpc$) path lengths \citep{SetaF2021pul} and might not matter much here, since we always integrate over path lengths greater than the scale of the large-scale field (a few $\kpc$). We further assume that there is a uniform density of $\RM$, but given a sufficient number of data points, the structure function analysis will not be affected significantly by this (\citealt{StutzkiEA1998}, also see Appendix~F in \citealt{SetaEA2023} for a numerical demonstration). Moreover, we have assumed that the emitted synchrotron emission is primarily Faraday rotated by the medium, but more generally, the synchrotron emission and Faraday rotation might occur together within the same volume of the galaxy \citep{VanEckEA2017, ThomsonEA2019, Ferriere2020}, leading to imprints of a {\it mixed and patchy signal} in $\RM$ observations. This might affect our results and we plan to characterise the level of this effect in our future studies.

Overall, the modelling used here allowed us to explore a large parameter space (\Tab{tab:notdef}) in a flexible way to devise new methods to extract large-scale magnetic field properties from second-order $\RM$ structure function in the presence of small-scale magnetic fluctuations, and we aim to include some of these additional effects in our future work.

\section{Summary and Conclusion} \label{sec:con}

Magnetic fields in disk galaxies are known to have a two-scale nature, a large-scale field (usually coherent over a few $\kpc$s) and a small-scale, fluctuating component (usually on $\lesssim 100$s of $\pc$ scales). From observations, it is usually difficult to separate the imprints of the large-scale component in the presence of small-scale fluctuations. Here, we use the second-order structure function (SF) of an observable, Faraday rotation measure ($\RM$, \Eq{eq:rm}), with higher-order stencils, to separate the signatures of small- and large-scale magnetic fields, and to determine their properties.

We employ numerical experiments with controlled properties of the magnetic field (both small- and large-scale with varying scale separation) and the thermal electron density, to develop a new method to extract large-scale field properties from the second-order RM structure function, $\RMSF$. We model the magnetic field by combining a single-scale, large-scale component (strength $\B$, characteristic wavenumber $\kB$, and related characteristic scale $\lB$) with a small-scale, Gaussian fluctuating component (strength $\brms$, characteristic wavenumber $\kbmin$, related characteristic scale $\lb$, and power-law spectrum $\Pb$) (see \Sec{sec:cons}). We then compute the $\sfn$, with a varying number of points per stencil, $\npts=2$ to $11$ (\Sec{sec:rmsf}), for a range of parameter combinations given in \Tab{tab:notdef}. Based on this, we now summarise the new method developed to extract the scale and strength of the large-scale magnetic field.

First, for a constant thermal electron density, we show that the $\RMSF$ reveals the two-scale nature if a sufficiently high number of points per stencil is used. We recover the input field structure, which is seen in the $\RMSF$ by a first rise (probe of the slope of the small-scale magnetic field power spectrum, discussed in \citet{SetaEA2023}), then saturation after scale $\lb/2$ to a level $\sigRMb$ (standard deviation of $\RM$ only considering the small-scale field), followed by another rise on larger scales within the range $\lb/2$ and $\lB/2$ (second rise), and finally saturation on large scales, greater than $\lB/2$ (giving the total $\sigRM$). This two-scale nature is better captured by $\RMSF$s with a higher number of points per stencil, in comparison to the widely-used 2-point-stencil case (\Fig{fig:allpts}).

The second rise in the $\RMSF$ can be fitted with a line in log-log space, and this line is extrapolated to $\sigRMb$ (the first saturation), which gives the intermediate scale $\ellLS$. This scale depends on the number of points per stencil and monotonically increases with $\npts$. In the theoretical limit of $\npts\to\infty$, the ratio $\LSrat\to1$, and in practice, using a couple $\npts$, $\lB$ can be determined from the universal functional form (\Fig{fig:scale}). Second, the ratio of $\sigRM$ (the second saturation level on larger scales) to $\sigRMb$ (the first saturation level on smaller scales) can be used to determine the strength of the large-scale magnetic field via the universal relation $(2 \kbmin L)^{-1}~(\rev{\sigRM^{2}/\sigRMb^{2}} - 1) \approx (\Brat)^{2}$ that we found in \Fig{fig:strength} (\Sec{sec:strength}). Thus, knowing $\lb/2$ (from the scale at which the first saturation takes place) and $\brms$ (from the level of first saturation, $\sigRMb$), one can determine $\B$. In \Sec{sec:ne}, we tested and determined that the large-scale properties of the $\RMSF$ are not significantly altered by using a log-normal thermal electron density distribution (as opposed to a constant value of $\ne$) of varying scale and slope of its power spectrum.

In \Sec{sec:lsobs}, we apply the developed methods to find the scale of the large-scale magnetic fields in the two nearby, face-on spiral galaxies, M51 and NGC~6946, by using RMSF data with 2~points per stencil from \citet{MaoEA2015} and \citet{Beck2007}, respectively. We find the large-scale field obtained in the disks of these galaxies ($\sim 30~\kpc$, see \Fig{fig:obs}) to be likely overestimated by a factor of $3$--$6$, and the large-scale field in the galaxy halos ($\sim 40$--$50~\kpc$; see \Fig{fig:obs}~(a)) are also likely overestimated. However, if the halo field has significant contributions from the circumgalactic medium of these galaxies, the scales may be in the expected range. We use these two galaxies as example applications to demonstrate the basic workings of the new method and to motivate the need for $\RMSF$s with higher-order stencils. We further discuss possible missing effects in \Sec{sec:mis}.

In summary, we developed new methods based on the second-order $\RM$ structure function computed with higher-order stencils that allow the user to recover key properties of multi-scale magnetic fields in galaxies, in particular, the scale and strength of the large-scale magnetic field in the presence of small-scale magnetic fluctuations.

\section*{Acknowledgements}
\rev{We thank Rainer Beck for very useful comments and suggestions on the paper. We further thank the anonymous referee for their fast and productive report.} AS acknowledges that this work has benefited from discussions during the program ``Towards a Comprehensive Model of the Galactic Magnetic Field'', at Nordita in April~2023, partly supported by the NordForsk and Royal Astronomical Society. C.~F.~acknowledges funding provided by the Australian Research Council (Discovery Project DP230102280), and the Australia-Germany Joint Research Cooperation Scheme (UA-DAAD). We further acknowledge high-performance computing resources provided by the Leibniz Rechenzentrum and the Gauss Centre for Supercomputing (grants~pr32lo, pr48pi and GCS Large-scale project~10391), the Australian National Computational Infrastructure (grant~ek9) and the Pawsey Supercomputing Centre (project~pawsey0810) in the framework of the National Computational Merit Allocation Scheme and the ANU Merit Allocation Scheme.

\section*{Data Availability}
The data underlying this article will be shared on reasonable request to the corresponding author, Amit Seta (\href{mailto:amit.seta@anu.adu.au}{amit.seta@anu.adu.au}). The data used in \Fig{fig:obs}~(a) and \Fig{fig:obs}~(b) was taken from \citet{MaoEA2015} and \citet{Beck2007}, respectively.

\bibliographystyle{mnras}
\bibliography{rmsflsf} 


\appendix

\bsp	
\label{lastpage}
\end{document}